\documentclass[11pt,letter]{article}

\newif\iffigures
\figurestrue 

\newif\ifproofs
\proofstrue

\usepackage{xcolor}
\usepackage{setspace}
\usepackage{amsmath} 
\usepackage{amssymb,nicefrac}
\usepackage[]{graphicx}
\usepackage{mathrsfs}

\newtheorem{definition}{Definition}[section]
\newtheorem{lemma}{Lemma}[section]
\newtheorem{theorem}{Theorem}[section]
\newtheorem{proposition}{Proposition}[section]

\newtheorem{assumption}{Assumption}[section]
\newenvironment{proof}{\textbf{Proof.}}{$\square$\\}

\definecolor{Yellow}{rgb}{1, 1, 0}
\definecolor{VeryLightGray}{gray}{.90}
\definecolor{LightGray}{gray}{.7}
\definecolor{Gray}{gray}{.50}
\definecolor{DarkGray}{gray}{.3}
\definecolor{VeryDarkGray}{gray}{.10}

\newcommand{\tr}{^{\mathrm T}}

\newcommand{\magn}[1]{\left\vert #1 \right\vert}

\newcommand{\vv}{v}
\newcommand{\vd}{d}
\newcommand{\tv}{v}

\newcommand{\cV}{\mathcal{V}}

\newcommand{\vs}{s}

\newcommand{\df}{\doteq}

\newcommand{\RM}{$\mathsf{RM}$}

\newcommand{\Simplex}[1]{\mathbf{\Delta}\left(#1\right)}
\newcommand{\EXP}[2]{\mathbb{E}_{#1}\left[#2\right]}
\newcommand{\Prob}[2]{\mathbb{P}_{#1}\left[#2\right]}
\newcommand{\ORDER}[1]{\mathcal{O}\left(#1\right)}
\newcommand{\DIST}[2]{{\rm dist}\left({#1},{#2}\right)}

\makeatletter
\newcommand{\xRightarrow}[2][]{\ext@arrow 0359\Rightarrowfill@{#1}{#2}}
\makeatother

\begin{document}

\title{Measurement-based Efficient Resource Allocation with Demand-Side Adjustments  \thanks{This work has been supported by the European Union grant EU H2020-ICT-2014-1 project RePhrase (No. 644235). It has also been partially supported by the Austrian Ministry for Transport, Innovation and Technology, the Federal Ministry of Science, Research and Economy, and the Province of Upper Austria in the frame of the COMET center SCCH. An earlier version of parts of this paper appeared in \cite{chasparis_reinforcement-learning-based_2015}.}}  
     
\author{Georgios C. Chasparis\thanks{G.~Chasparis is with the Department of
Data Analysis Systems, Software Competence Center Hagenberg GmbH, Softwarepark 21, A-4232 Hagenberg, Austria; E-mail: {\rm georgios.chasparis@scch.at}.} }



\date{December 18, 2017 \\ \today \ \ (revised) }

\maketitle

\begin{abstract}
The problem of efficient resource allocation has drawn significant attention in many scientific disciplines due to its direct societal benefits, such as energy savings. Traditional approaches in addressing online resource allocation problems neglect the potential benefit of feedback information available from the running tasks/loads as well as the potential flexibility of a task to adjust its operation/service-level in order to increase efficiency. The present paper builds upon recent developments in the area of bandwidth allocation in computing systems and proposes a generalized design approach for resource allocation when only performance measurements of the running tasks are available, possibly corrupted by noise. We demonstrate through analysis and simulations the potential of the proposed scheme in providing fair and efficient allocation of resources in a large class of resource allocation problems.
\end{abstract}

\section{Introduction}
Resource allocation has become an indispensable part of the design of many engineering systems that consume resources, such as electricity power in home energy management \cite{DeAngelis13}, access bandwidth and battery life in wireless communications \cite{Inaltekin05}, computing bandwidth and memory in parallelized applications \cite{Brecht93}, computing bandwidth in CPU cores \cite{Bin11}. 

When resource allocation is performed online and the number, arrival and departure times of the tasks are not known a priori (as in the case of CPU bandwidth allocation), the role of a resource manager (\RM) is to guarantee an \emph{efficient} operation of all tasks by appropriately distributing resources to the tasks and also assigning their operation/service-levels. However, guaranteeing efficiency through the adjustment of resources and/or operation-levels requires the formulation of a centralized optimization problem (e.g., mixed-integer linear programming formulations \cite{Bin11}), which further requires information about the specifics of each task and their response under different resource/operation-level pairs. Such information may not be available to neither the \RM\ nor the task itself.

Given the difficulties involved in the formulation of centralized optimization problems, not to mention their computational complexity, feedback from the running tasks in the form of performance measurements may provide valuable information for the establishment of efficient allocations. Such (feedback-based) techniques have recently been considered in several scientific domains, such as in the case of application parallelization (in the form of scheduling hints) \cite{Broquedis10}, or in the case of allocating virtual platforms to computing applications \cite{Ste99}. Recently, a measurement-based learning scheme has been proposed \cite{chasparis_design_2016} specifically tailored to the problem of CPU-bandwidth allocation for time-sensitive applications. This scheme exhibits the benefits of measurement- or feedback-based methods, while, in parallel, allows  applications for adjusting their own operation-level. 

Motivated by the framework of \cite{chasparis_design_2016} and the potential of exploiting both measurements available from the tasks and the flexibility of some tasks in changing their operation-level, in this paper we propose a generalized design methodology for addressing a general class of online resource allocation problems. In particular, the \RM\ is responsible for adjusting both the resources and the operation-levels of the task, where the adjustment processes are based only on performance measurements received from the running tasks. The proposed scheme exhibits adaptivity and robustness in the number, type and performance variations of the tasks. We demonstrate through analysis the potential of the proposed scheme in the establishment of fair and efficient allocations for a large class of resource allocation problems.

The paper is organized as follows. Section~\ref{sec:RelatedWork} discusses related work and the main contributions. Section~\ref{sec:framework} formulates a centralized optimization problem for a general class of online resource allocation problems and provides two examples. Section~\ref{sec:LearningDynamics} presents a learning scheme for adjusting both resources and operation-levels of the tasks. Section~\ref{sec:ResourceLevelConvergenceProperties} presents convergence properties of the resource adjustment, while Section \ref{sec:OverallConvergenceProperties} presents convergence properties of the combined resource and operation-level adjustment. Section~\ref{sec:Simulations} provides a simulation study in the context of power management in residential buildings. Finally, Section \ref{sec:Conclusions} presents concluding remarks.

\textit{Notation:} 
\begin{itemize}
\item $\Pi_{[a,b]}$ is the projection onto the set $[a,b]$.
\item For any $x\in\mathbb{R}^{n}$ and set $A\subset{\mathbb{R}^{n}}$, define $\DIST{x}{A}\df\inf_{y\in{A}}|x-y|,$ where $|\cdot|$ denotes the Euclidean norm.
\item For some set $A\subset\mathbb{R}^{n}$ and $\delta>0$, define its $\delta$-neighborhood as $\mathcal{B}_{\delta}(A) \df \left\{x\in\mathbb{R}^{n}: \DIST{x}{A}\leq\delta\right\}.$
\item The probability simplex of dimension $n$ is defined as 
$\Simplex{n}=\{x=(x_1,...,x_n)\in[0,1]^n:\sum_{i=1}^nx_i = 1\}.$
\item For some finite set $A$, $\magn{A}$ denotes the cardinality of $A$.
\item For any matrix $A\in\mathbb{R}^{m\times{n}}$, $A\tr$ denotes its transpose.
\end{itemize}

\section{Related Work and Contributions}
\label{sec:RelatedWork}

Efficiency in resource allocation problems has been addressed in several scientific domains under different sets of assumptions. A popular approach in addressing efficiency in multi-agent/multi-component systems is through \emph{subgradient optimization} of an overall welfare criterion \cite{nedic_distributed_2009,johansson_randomized_2009}. Under such schemes, and in the context of resource allocation, a group of tasks \emph{cooperatively} try to maximize an overall welfare criterion of the form $\sum_{i=1}^{n}u_i(v)$, where $u_i$ represents the performance index of task $i$ under the provided resources $v$. A common assumption under subgradient optimization schemes is the (a-priori) knowledge of the details of the performance functions $u_i$, including additional structural assumptions, such as convexity (see, e.g., \cite{nedic_distributed_2009}). Related methods that drop the convexity structural assumption are presented in \cite{zhu_approximate_2013,tatarenko_non-convex_2017} for convergence to local optima. In the presence of feasibility constraints, as it is usually the case in the context of resource allocation problems, a gradient-based algorithm is presented in \cite{yi_initialization-free_2016}, where constraints are only satisfied asymptotically (\emph{soft constraints}).

In several practical scenarios, a-priori knowledge of the partial derivatives of the performance functions might be quite restrictive or practically impossible. Furthermore, the structural assumption of convex performance functions may also be restrictive. Even if convexity is not required, as in \cite{zhu_approximate_2013,tatarenko_non-convex_2017}, guaranteeing \emph{efficiency} (i.e., maximum overall welfare) does \emph{not} necessarily guarantee \emph{fairness}. For example, maximizing social welfare $\sum_{i=1}^{n}u_i(v)$ will not necessarily guarantee that all tasks receive non-zero amount of resources.

Parallel to this line of research in subgradient optimization schemes, several resource allocation problems have also been addressed within the context of game-theoretic methods. The main goal of such approaches is that the solution of a centralized (global) optimization problem is addressed through agent-based (local) objectives, where agents may represent the tasks to be allocated. Examples include the cooperative game formulation for allocating bandwidth in grid computing \cite{Sub08}, the non-cooperative game formulation in the problem of medium access protocols in communications \cite{Tembine09} or for allocating resources in cloud computing \cite{Wei10}. When allocation decisions belong to a finite set (e.g., when agent/task $i$ may select a subset of a finite set of resources) resource allocation can be addressed within the context of \emph{submodular} optimization problems, as in \cite{marden_overcoming_2009}, and \emph{coordination games}, as in \cite{ChasparisAriShamma13_SIAM}. Some examples include common-pool problems \cite{Meinhardt99,ChasparisAriShamma13_SIAM}, bin-packing problems \cite{epstein_selfish_2008} and load-balancing problems \cite{vocking_selfish_2007}. In the presence of soft constraints, relevant payoff-based learning may also include the learning-automata dynamics for convergence to Nash equilibria in convex games of \cite{tatarenko_learning_2019}.

There are three main difficulties involved in the implementation of the aforementioned game-theoretic schemes in practical scenarios. First, the designer may not have full control over the properties of the performance indices of the tasks. For example, if a task represents a computing application, the performance index may be a function of the processing speed, however we do not know how the processing speed varies with the provided resources. Second, when resources are limited and should be shared between tasks, hard constraints may need to apply at all times (and not only asymptotically). Third, although convergence to a Nash equilibrium or Pareto efficient outcomes can be guaranteed under several classes of payoff- or measurement-based learning (as in the benchmark-based dynamics of \cite{marden_pareto_2014} or the reinforcement-learning dynamics of \cite{chasparis_stochastic_2019}), \emph{fairness} can be guaranteed only under special classes of games (as in the common-pool games of \cite{ChasparisAriShamma13_SIAM}).

Contrary to the aforementioned literature, we wish to address a class of resource allocation problems where a pool of resources need to be \emph{fairly} allocated to a set of running tasks. Each task receives a portion of the available resources, i.e., tasks evolve over the probability simplex.  Naturally, in these problems \emph{feasibility} of the allocation constitutes a hard constraint, thus imposing the presence of a \RM. The specific contributions are:

\begin{enumerate}

\item We introduce a generic \emph{fairness} measure, which, in the context of measurement-based optimization, can guarantee a) \emph{feasibility}, b) \emph{starvation avoidance}, and c) \emph{balanced} allocations. These results require only continuity of the performance function with respect to the provided resources. In other words, we do not impose any strong structural assumptions in the interdependence of performances between the running tasks (as in the aforementioned game-theoretic schemes). We also address \emph{fairness} in the allocation of resources, contrary to subgradient-based optimization schemes or game-theoretic methods mentioned above.

\item No a-priori knowledge of the performance functions of the tasks is assumed, contrary to standard subgradient-based schemes. In fact, only measurements of their performance indices are considered, possibly corrupted by noise.

\item An orthogonal dimension of optimization is also considered, where the specifications or operation-level of a given task can also be adjusted to further accommodate flexibility in the adjustment of resources. To the best of our knowledge, such additional dimension of optimization is not considered in prior resource allocation schemes.

\end{enumerate}

The above contributions extend prior work of the author \cite{chasparis_reinforcement-learning-based_2015,chasparis_design_2016}, in (1) generalizing the definition of fairness to a generic class of problems (not only restricted to real-time computing applications), (2) incorporating noise in the performance measurements, and (3) providing global convergence guarantees under synchronous resource and operation-level updates.

\section{Problem Formulation and Objective} 
\label{sec:framework}

\subsection{Framework}
\label{sec:RMapp}

We consider a resource allocation framework where one or multiple users request a finite number of tasks $\mathcal{I}=\{1,2,...,n\}$ to be executed. We denote such requests by $d_i\in\mathcal{D}_i$, indicating the \emph{demand} of a user with respect to task $i$. Each of these tasks may run at a different \emph{operation-} or \emph{service-level}, denoted by $s_i\in\mathcal{S}_i$, indicating the level of comfort provided to the user through task $i$. We admit a normalization of the space of operation-levels, i.e., we consider $\mathcal{S}_i\df[0,1]$, ranging between its two extreme values.

In order for a task to be executed, an amount of resources needs to be assigned to it which corresponds to a portion $v_i\in\cV_i\df[0,1]$ of the overall available resource. In other words, $\tv_i$ corresponds to the rate of accessing a common good. Here it is implicitly assumed that there is one type of available resource. Examples include a) electrical power in residential buildings where tasks represent electrical loads, b) computing bandwidth in CPU cores, and c) bandwidth in wireless communications. 

\begin{figure}[t!]
\centering
\includegraphics[scale=1]{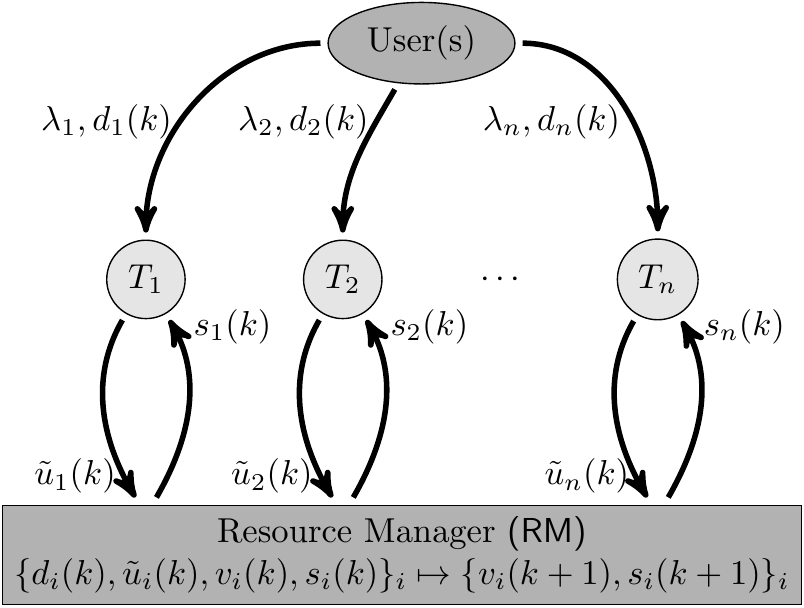}
\caption{Schematic of resource allocation framework.}
\label{fig:framework}
\end{figure}

The operation-level of each task, $s_i$, and the amount of resources assigned to it, $v_i$, are determined by a \emph{resource manager} (\RM) which is responsible for maintaing a desirable performance of the overall system (according to some user-defined criterion). The \RM{} makes decisions about the resources and service levels of the tasks at regular time instances denoted by $k=0,1,2,...$. The assignment of resources and service levels to the tasks is based solely on performance measurements received from each task, denoted by $\tilde{u}_i$. Throughout the paper, we consider the following assumption.
\begin{assumption}
The \RM{} satisfies the following design properties:
\begin{itemize}
\item (D1) The internal characteristics of the tasks are not known to the \RM{}. Instead, the \RM\ may only have access to measurements related to their performance. 
\item (D2) Tasks may not be split, rescheduled or postponed. Instead, the goal of the \RM{} is to assign the currently available resources to the currently requested tasks (work-conserving). 
\end{itemize}
\end{assumption}

Note that the design assumptions (D1) and (D2) describe a framework in which the starting time of a task is \emph{not} an optimization parameter. The \RM\ does not have the necessary information to make such scheduling decisions (e.g., it does not know the duration time of a task). Thus, the main question is how to efficiently assign resources to the tasks assuming that they should immediately start running upon creation. 

\begin{figure}[t!]
\centering
\includegraphics[scale=1]{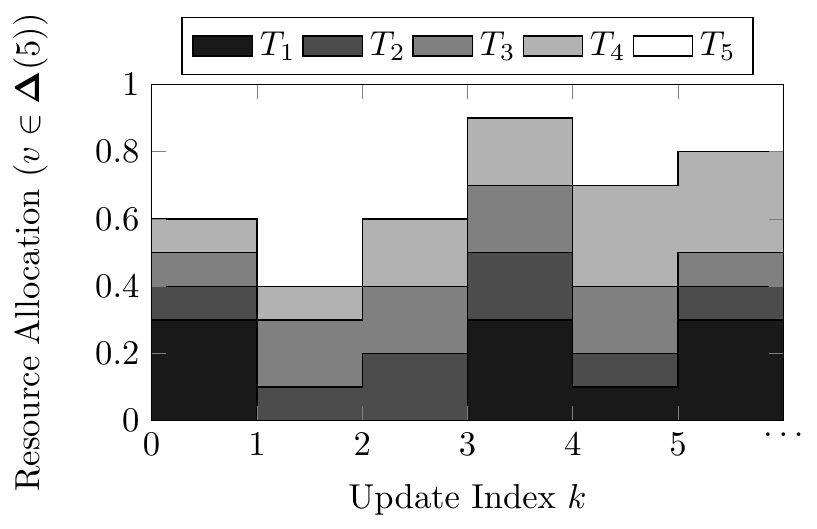}
\caption{Schematic of resource allocation evolution for five tasks $i=1,2,...,5$, denoted by $T_1$, $T_2$,...,$T_5$, respectively.}
\label{fig:allocation}
\end{figure}

\begin{figure}[t!]
\centering
\includegraphics[scale=1]{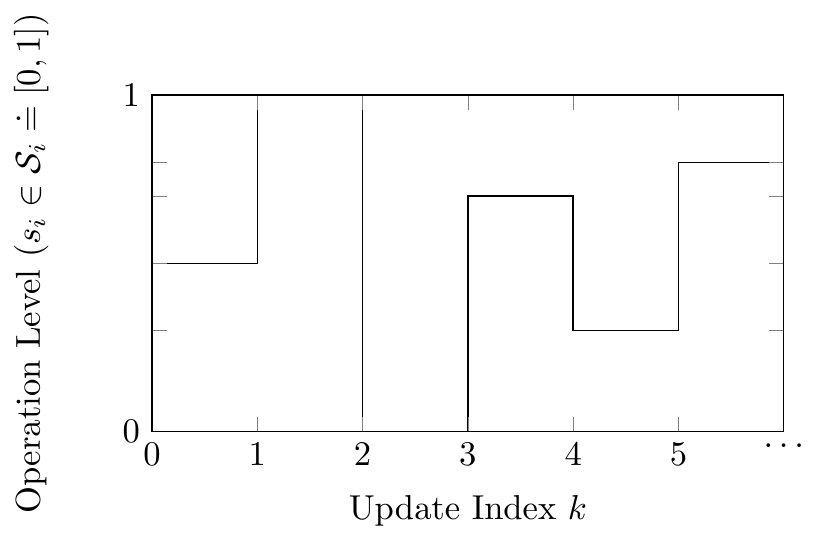}
\caption{Schematic of operation-level evolution for some task $i$.}
\label{fig:servicelevel}
\end{figure}

The overall framework is illustrated in Figure~\ref{fig:framework} describing the flow of information, starting from the users who determine the requests and ending to the \RM{} which recursively allocates resources $\vv=(v_1,...,v_n)$ and operation-levels $\vs=(s_1,...,s_n)$ to the tasks in $\mathcal{I}$ based on the collected measurements. Figure~\ref{fig:allocation} demonstrates schematically how an allocation of resources $\vv$ may look like for a set of tasks requesting resources at regular time intervals. We assume that allocations belong to the set $\cV\df\Simplex{n}$, since each resource $v_i$ may only be a portion of the total available resource corresponding to 1. Figure~\ref{fig:servicelevel} demonstrates how the operation-level $s_i\in\mathcal{S}_i$ of a task $i$ may evolve as updated by the \RM{} over time. We will also use the notation $\mathcal{S}\df\mathcal{S}_1\times\ldots\times\mathcal{S}_n$ and $\mathcal{D}\df\mathcal{D}_1\times\ldots\times\mathcal{D}_n$ which are the domains of operation-level and demand profiles, respectively.


Note that \emph{the amount of resources assigned by the \RM{} to the currently requested tasks may not necessarily correspond to the amount of resources used by each task}. The amount of resources used by each task depend solely on the operation-level $s_i$ and the corresponding demand $d_i$. Informally, we may say that the resource allocation $\vv$ constitutes a form of recommendation provided by the \RM. Whether this recommendation is indeed implemented depends on whether the operation-level profile is appropriately set to use efficiently the recommended amount of resources. These points will become more obvious shortly when we discuss these terms through some application scenarios.

\subsection{Utility function and efficiency}	\label{sec:UtilityFunction}

The objective is two-fold. On the one hand, the \RM{} is responsible for maintaining a \emph{fair} allocation, $\vv$, among the requested tasks, while, on the other hand, the service level of each task $i$ should guarantee an \emph{efficient} operation given the amount of resources $v_i$ provided by the \RM{}. Before introducing the notions of \emph{fairness} and \emph{efficiency}, we first need to introduce the \emph{performace measure} or \emph{utility function} for each task $i$.

\subsubsection{Utility Function}
\label{sec:UtilityFunction}

The utility of a task $i$ is introduced to capture the fitness of the task conditional to the amount of resources $v_i$ provided by the \RM, its operation-level $s_i$ and the user demand $d_i$. It is defined as a function of the form $u_i:\mathcal{S}_i\times\mathcal{V}_i\times\mathcal{D}_i\mapsto\mathbb{R}_+$, where we employ the following conditions.
\begin{assumption}[Utility function]  \label{As:UtilityFunction}
The utility function $u_i:\mathcal{S}_i\times\mathcal{V}_i\times\mathcal{D}_i\mapsto\mathbb{R}_+$ of a task $i\in\mathcal{I}$ is continuous with respect to its arguments and satisfies:
\begin{itemize}
\item (U1) There exists a positive constant $c_i>1$, such that $$1 \leq u_i(s_i,v_i,d_i)<c_i$$ uniformly on $v_i\in\mathcal{V}_i$, $s_i\in\mathcal{S}_i$ and $d_i\in\mathcal{D}_i$.
\item (U2) For any given allocation $v_i$ and demand $d_i$, $u_i(\cdot,v_i,d_i)$ is continuously differentiable with respect to its first argument $s_i\in\mathcal{S}_i$ and concave. 
\end{itemize}
\end{assumption}

Note that these are design assumptions that may be used to well represent the reasoning of a performance index in a resource allocation problem. In particular, we should always expect (given the boundedness of the provided resources) that (U1) the utility function is uniformly bounded from above. The condition of the utility function being greater than the unity is introduced for technical reasons and can always be met by appropriately shifting the utility function.  
Finally, we should further expect that as the operation-level increases, then (U2) the corresponding performance should increase, however the gradient of the performance should saturate given the limited amount of resources.

\subsubsection{Fairness and efficiency}	\label{sec:Efficiency}

Assuming that the utility function for each task has been designed, we introduce the following \emph{\textbf{fairness measure}}:
 \begin{eqnarray*} \label{eq:FairAllocationMeasure} 
  \lefteqn{{\Phi}_i(\vs,\vv,\vd) \df}\cr && (1-v_i)\lambda_i [u_i(s_i,v_i,d_i)]^{-1} - v_i \sum_{j\neq{i}}\lambda_j [u_j(s_j,v_j,d_j)]^{-1},
\end{eqnarray*}
for some constants, $\lambda_i\in(0,1]$, $i\in\mathcal{I}$.

The function $\Phi_i$ \emph{captures the deficiency in resources of task $i$ as compared to the rest of the tasks}. When task $i$ is not performing well in comparison with the rest of tasks, i.e., $[u_i(s_i,v_i,d_i)]^{-1}$ is significantly larger than $[u_j(s_j,v_j,d_j)]^{-1}$, $j\neq{i}$, and its available resources $v_i$ are small, we should expect that $\Phi_i$ admits large (positive) values (indicating deficiency of resources for task $i$). If, instead, task $i$ is performing well, while it also has large amount of resources $v_i$, then we should expect that $\Phi_i$ admits small (negative) values (indicating sufficiency of resources for task $i$). The factor $\lambda_i\in[0,1]$ which scales the inverse utility represents the importance of the task and it is user-defined. 

\begin{definition}[Fair and efficient allocation] \label{def:EfficientAllocation} 
For some given demand profile $\vd=(d_1,...,d_n)\in\mathcal{D}$, an allocation of resources $\vv^*\in\cV$ and operation-levels $\vs^*\in\mathcal{S}$ is called fair if $\Phi_i(s^*,v^*,\vd)\equiv{0}$ for all tasks $i$, and efficient if $u_i(s_i^*,v_i^*,d_i)\to\max$ for all tasks $i$.
\end{definition}

We will often denote by $\mathcal{F}^*=\mathcal{F}^*(d)$ and $\mathcal{E}^*=\mathcal{E}^*(d)$ the set of fair and efficient allocations, respectively. Thus, a pair $(\vv^*,\vs^*)\in\mathcal{F}^*\cap\mathcal{E}^*$ will provide an ideal operation with respect to both a) the allocation of resources, and b) the operation of each task separately.

First, note that, according to the introduced fairness measure, 
\begin{equation}
\mbox{ if } \tv_i=0, \mbox{ then } \Phi_i = \lambda_i [u_i(s_i,\tv_i,d_i)]^{-1} > 0
\end{equation}
which implies that $\tv_i>0$ is a necessary condition of any fair allocation. Furthermore, according to Definition~\ref{def:EfficientAllocation}, an allocation $\vv^*$ is \emph{fair} if and only if the provided resources are ``balanced'' with the corresponding performances. To see this, note that fairness implies: 
$$\frac{\tv_i^*}{\sum_{j\neq{i}}\tv_j^*} = \frac{\lambda_i[u_i(s_i,\tv_i^*,d_i]^{-1}}{\sum_{j\neq{i}}\lambda_j[u_j(s_j,v_j^*,d_j)]^{-1}},$$ given that $1-v_i^*=\sum_{j\neq{i}}v_j^*$. To understand this identity, let us consider the case of equal weights, i.e., $\lambda_1=...=\lambda_n$. If $v_i^*$ is large as compared to the rest of the resources, $\sum_{j\neq{i}}v_j^*$, then $[u_i]^{-1}$ has to be sufficiently large, i.e., task $i$ should not perform so well in comparison with the rest of the tasks. Informally, there could \emph{not} be a task $i$ that monopolizes the resources at a fair allocation when $i$ performs well and the others do not. Large amount of resources in a single task may only be justified if the task is performing poorly in comparison with the rest of the tasks. If we allow for non-uniform weights $\lambda_i$, then large amount of resources in a single task may also be justified by a large weight. In the trivial case of \emph{identical} tasks, with $s_1=...= s_n$, $d_1=...= d_n$ and $\lambda_1=...=\lambda_n$, one allocation that satisfies the fairness condition is $\tv_i^* = \nicefrac{1}{n}.$

A similar fairness measure has been introduced within the context of CPU Bandwidth Allocation problem in \cite{chasparis_design_2016}. The above definition is more general since the performance indices are not restricted to any specific application scenario and they are only subject to Assumption~\ref{As:UtilityFunction}. 

\subsection{Examples}	\label{sec:Examples}

To demonstrate the utility of the proposed framework, let us discuss the following practical scenarios.

\subsubsection{Home Energy Management}	\label{sec:ExampleHomeEnergyManagement}

A simplified version of the smart-home paradigm considers a central \RM\ which controls the amount of electrical power assigned to the electricity loads demanded by the user. In this case, $v_i\in[0,1]$ represents the power assigned to each load where the maximum value 1 corresponds to the (desirable) maximum power available. Note that this might \emph{not} be the actual power used.

Loads may correspond to flexible loads, such as the operation of the heating system, heat pumps or lighting. The user may define set-point temperatures for the operation of the heating system and the heat pumps, and desired luminance levels for lighting. Such set points may be considered as demand requirements, $d_i$. 

The operation-level $s_i$ of each task $i$ may correspond to the different levels of the service provided. For example, in the case of the heating system, it may correspond to the heating input provided to each thermal zone of the building, while in the case of the lighting equipment, it may represent the luminance level provided to each zone. 

The definition of the utility function that may represent the operation of these tasks is open-ended. Consider the trivial example of the greedy objective for maximizing the comfort level provided by each task, which may be represented by a utility function of the form:
\begin{equation}	\label{eq:UtilityHomeEnergyManagement}
u_i(s_i,v_i,d_i) \triangleq a \ell_i(s_i,d_i) + b (v_i - e_i(s_i)) + c,
\end{equation}
for some positive constants $a$, $b$ and $c$, where the function $\ell_i$ captures the comfort of the user, while the function $e_i$ corresponds to the energy rate consumed by task $i$. Note that any excess energy rate from the assigned $v_i$ (i.e., when $v_i<e_i(s_i)$) is penalized, while any energy rate savings (i.e., when $v_i>e_i(s_i)$) is encouraged. Alternative functions can be defined depending on the application and the performance indices which can be measured. The parameters of such objective function may be user-defined.

Let us consider the example of the heating system in a residential building as described in detail in \cite{chasparis_regression_2016}. In this example, the comfort of the user can be described as $\ell_i(s_i,d_i)\df \kappa - (s_i-d_i)^2,$ for some positive constant $\kappa>0$. The comfort admits its maximum value $\kappa$ when the operation-level meets the corresponding demand, i.e., $s_i\equiv d_i$. In any other case, the comfort admits lower values than $\kappa$. Furthermore, the heating cost of a radiant heating system can be approximated by a linear function of the flow rate of the thermal medium (in this case, the operation-level), i.e., $e_i(s_i) \df h s_i,$ for some $h>0$. Thus, in the case of the heating system, the utility function of the task takes on the form:
\begin{equation*}
u_i(s_i,v_i,d_i) = a\left(\kappa - (s_i-d_i)^2\right) + b(v_i- h s_i) + c,
\end{equation*}
for some positive constants $a$, $b$ and $c$ such that condition (U1) is satisfied. It is also straightforward to verify that the utility function is continuous with respect to its arguments and that (U2) the utility function is concave with respect to $s_i$, since $\nabla_{s_i}^2 u_i(s_i,v_i,d_i) = -2a < 0$.

\subsubsection{CPU Bandwidth Allocation}	\label{sec:ExampleCPUBandwidthAllocation}

The above framework may accommodate resource allocation problems encountered in the context of CPU bandwidth allocation. Recent work \cite{chasparis_design_2016} has focused on designing such utility functions for the case of time-sensitive applications. In this scenario, the \RM{} is responsible for assigning \emph{virtual-platforms} $v_i$ to each application $i$. We may think of $v_i$ as the percentage/portion of the CPU assigned to application $i$, which determines the rate with which an application $i$ executes a job and the corresponding time interval assigned to the application. 

Specifically in the case of time sensitive applications, including for example multimedia and control applications, the performance of the application depends on the relation between the \emph{response-time} of a job $R_i$ and the corresponding \emph{soft-deadline} for executing a job, $D_i$, (determined by $v_i$). Good performance translates to $R_i\equiv D_i$. A natural definition of such a performance function may take on the following form,
\begin{equation}	\label{eq:UtilityCPUBandwidthManagement}
u_i(s_i,v_i,d_i) \df -a (D_i(d_i,s_i) - R_i(s_i,v_i))^2 + b,
\end{equation}
for some constants $a,b>0$ selected appropriately so that condition (U1) is satisfied. Note that the utility function attains a unique maximum when the deadline $D_i$ approaches $R_i$, which is the desired property.

As described in \cite{chasparis_design_2016}, and in the context of multimedia applications, the soft deadline $D_i$ can be considered constant, e.g., $D_i=h>0$, while the response time can be defined as $R_i=\nicefrac{C_i}{v_i}$, where $C_i=\theta_is_i$ is the execution time per job (at a service-level $s_i$), for some $\theta_i>0$, and $v_i$ is the speed of execution. In this case, the utility of application $i$ takes on the following form:
\begin{equation}
u_i(s_i,v_i,d_i) = - a\Big(h-\theta_i\frac{s_i}{v_i}\Big)^2 + b.
\end{equation}
It is straightforward to check that this function is continuous with respect to $\tv_i$. Furthermore, it is concave with respect to $s_i$, since $\nabla^{2}_{s_i}u_i(s_i,v_i,d_i) = -2a(\theta_i/v_i)^2 < 0$.

\subsection{Objective}	\label{sec:Objective}

Ideally, we would like to set up a centralized optimization problem, solved by the \RM, such that at each update instance $k$, it would assign resources in an \textit{efficient} manner to all tasks. Definition~\ref{def:EfficientAllocation} introduces a potential centralized problem for efficient allocations, a candidate form of which is:
\begin{eqnarray} \label{eq:CentralizedObjective}
\centering
\begin{array}{ll}
  \min_{\vs\in\mathcal{S},\vv\in\cV} & \sum_{i\in\mathcal{I}}\magn{\Phi_i(\vs,\vv,\vd)}^2 \cr
  \mbox{s.t.} & s_i = \arg\max_{s\in\mathcal{S}_i}u_i(s,v_i,d_i), \quad i\in\mathcal{I},
\end{array}
\end{eqnarray}
for some given $d_i\in\mathcal{D}_i,$ $i\in\mathcal{I}.$

Whether such an optimization problem is well posed and the type of solutions it may accept depend on the characteristics of the utility functions $u_i$. Our goal is not to address directly such centralized optimization problem. This is because the definition of the utility function will necessarily be based upon measurements of quantities related to the performance of a task, whose explicit relation to the (internal) variables $s_i$ and the provided resources $v_i$ is \emph{not} known in general. 

To see this, let us consider the example of home energy management discussed in Section~\ref{sec:ExampleHomeEnergyManagement}. Note that the function $e_i(s_i)$ captures the energy consumed by the task. It can be measured, however its explicit relation to the operation-level $s_i$ is not known a-priori to the \RM{} (i.e., the parameter $h$ is unknown). Similar is also the case in the CPU Bandwidth Allocation problem, where the deadline $D_i$ and the response-time $R_i$ can be measured by the \RM{}, however their explicit dependencies on the resource level $v_i$ and operation-level $s_i$ are not known.

The \RM{} may only respond to measurements available, and thus addressing a centralized optimization problem as stated above is \emph{not} possible. \emph{The goal of this paper is to investigate a class of utility- or measurement-based learning dynamics in addressing computation of fair/efficient pairs $(\vs^*,\vv^*)$ as defined in Definition~\ref{def:EfficientAllocation}.}

\section{Learning Dynamics}
\label{sec:LearningDynamics}

Given the difficulties in formulating centralized optimization problems in the absence of explicit knowledge of the characteristics of the tasks requesting resources, we propose an adaptive scheme which is based on \emph{learning-based} (or \emph{measurement-based}) dynamics. According to the proposed scheme, the \RM\ is responsible for updating both the resource allocation $\vv$ and the operation-levels $\vs$ of the tasks. The goal is to attain convergence to an efficient allocation when only measurements of the utility functions are provided. 

\subsection{Resources update}
\label{sec:LearningDynamicsRM}

At time instances $t_k$, indexed by $k=0,1,\ldots$, the \RM\ measures the utility function of each task $i\in\mathcal{I}$ and updates the resources assigned to $i$ as follows:\footnote{We have intentionally omitted the constraint $\tv(k)\in\cV$, since it is always satisfied when the step-size $\epsilon$ is sufficiently small (as it will become evident by the forthcoming Proposition~\ref{Pr:Feasibility}).}
\begin{equation}	\label{eq:RecursionForResources}
  v_i(k+1) = v_i(k) + \epsilon F_i(k,v_i(k)),
\end{equation}
for each $i=1,...,n$, where $F_i$ is the \emph{observed} fairness index defined as follows: 
\begin{eqnarray*}  
\lefteqn{F_i(k,\tv_i(k)) \df } \cr &&  (1-v_i(k))\lambda_i [\tilde{u}_i(k)]^{-1} - v_i(k) \sum_{j\neq{i}}\lambda_j [\tilde{u}_j(k)]^{-1}.
\end{eqnarray*}
The quantity $\tilde{u}_i(k)$ denotes the measurement of the utility function of task $i$ which admits the form:
\begin{equation}	\label{eq:NoisyObservations}
\tilde{u}_i(k) = u_i(s_i(k),v_i(k),d_i(k)) + \eta_i(k),
\end{equation}
where $\eta_i(k)$ is a zero-mean bounded measurement noise, i.e., $\sup_{i\in\mathcal{I}}\magn{\eta_i}\leq\overline{\eta}$ for some $\overline{\eta}>0$. We further assume that this noise process is independently distributed for each $i\in\mathcal{I}$. The introduction of the noise process is necessary in order to capture some irregularities of the tasks (e.g., processes in computing systems). However, the type of the noise process cannot be known a-priori. The boundedness assumption is an indirect implication of the nature of the problems considered here, given that the performance indicators cannot deviate significantly from a nominal value (e.g., energy, time response, processing speed, etc.). However, even in the case that large noise values can be justified, for security reasons lower and upper bounds should be artificially introduced in all measured quantities.

According to the definition of $F_i$, if there is a deficiency of resources for $i$, i.e., $F_i>0$, then $v_i$ will increase, otherwise it will decrease. We consider a \emph{constant} step-size $\epsilon>0$, since it provides an adaptive response to changes in the number of applications.

The above recursion (\ref{eq:RecursionForResources}) extends prior work of the author where $F_i$ was specifically tailored to time-sensitive computing applications \cite{chasparis_design_2016}. The learning framework proposed here is independent of the nature of the tasks, as discussed in Section~\ref{sec:Examples}, while it also incorporates possibly corrupted observations.

\subsection{Operation-level update}
\label{sec:AppAdjust}

Due to the concavity of the utility function $u_i$ with respect to the operation-level $s_i$, a gradient-based learning dynamics can be introduced for updating the operation-level $s_i$, for each task $i$. Similarly to the case of the resource update, the explicit form of the utility function may not be known to the \RM\, thus we may only make use of measurements of the utility function. 

We would like that the operation-level updates take place at a faster timescale as compared to the resource update (\ref{eq:RecursionForResources}). The reason for this choice is the better control over the resulting convergence properties of the overall dynamics, since any decision over the allocation of resources will be performed with the operation-level updates being nearly equilibrated. The introduced faster timescale of the operation-level updates is also supported by the fact that can be locally implemented by each task, contrary to the resource updates that incur computational overhead in the \RM.

To this end, we introduce the following recursion for the operation-level of each task $i$.
\begin{eqnarray}	\label{eq:RecursionForOperationLevels}
\lefteqn{s_i(k+1) = } \cr 
&& \Pi_{[0,1]}\left[ s_i(k) + \epsilon \mu(\epsilon) \tanh\left(\frac{\tilde{U}_i(k)}{\tilde{S}_i(k)}\right) + \epsilon\mu(\epsilon) \zeta_i(k)\right]
\end{eqnarray}
where $\mu(\epsilon)$ is defined so that 
\begin{equation}	\label{eq:StepSizeConstraints}
\lim_{\epsilon\downarrow{0}}\epsilon \mu(\epsilon)={0}, \quad \lim_{\epsilon\downarrow{0}}\frac{\epsilon}{\epsilon \mu(\epsilon)}={0},
\end{equation} 
i.e., $\epsilon$ goes faster to zero than $\epsilon \mu(\epsilon)$, as $\epsilon\downarrow{0}$. Thus, the update recursion (\ref{eq:RecursionForOperationLevels}) moves on a faster timescale than recursion (\ref{eq:RecursionForResources}). The term $\zeta_i(k)$ corresponds to an artificially introduced zero-mean bounded noise term defined as $\zeta_i(k) \df {\rm rand}([-\overline{\zeta},\overline{\zeta}])$, for some positive constant $\overline{\zeta}>0$. The quantities $\tilde{U}_i(k)$ and $\tilde{S}_i(k)$ are approximations of the gradient of the measured performance $\tilde{u}_i(k)$ and the operation-level $s_i(k)$, respectively. They can be generated as low-pass filters of the measured quantities (motivated by \cite{shamma_dynamic_2005}), as follows
\begin{eqnarray*}
\tilde{U}_i(k) & \df & \gamma \cdot (\tilde{u}_i(k) - \rho_i(k)) \\
\tilde{S}_i(k) & \df & \gamma \cdot (s_i(k) - \sigma_i(k))
\end{eqnarray*}
for some $\gamma>0$, where $\rho_i(k)$ and $\sigma_i(k)$ are 
\begin{subequations}	\label{eq:RateApproximations}
\begin{eqnarray}
\rho_i(k+1) & = & \rho_i(k) + \epsilon \mu(\epsilon) \cdot \tilde{U}_i(k) \\
\sigma_i(k+1) & = & \sigma_i(k) + \epsilon \mu(\epsilon) \cdot \tilde{S}_i(k).
\end{eqnarray}
\end{subequations}

Note that the higher the value of $\gamma>0$, the better the approximation of the gradients. Thus, as $\gamma$ increases, we should expect that $s_i$ changes in the direction of increasing the utility $\tilde{u}_i$. This will formally be explained when we discuss the convergence properties of the overall recursion in the forthcoming Section~\ref{sec:OverallConvergenceProperties}.

\subsection{Overall update recursion}

It will be helpful to analyze the overall recursion dynamics as a whole, leading to the following set of recursions
\begin{eqnarray}	\label{eq:OverallUpdateRecursion}
\lefteqn{
\left(\begin{array}{c}
\tv_i \\
s_i \\
\rho_i \\
\sigma_i
\end{array}\right)(k+1) = 
\left(\begin{array}{c}
\tv_i \\
s_i \\
\rho_i \\
\sigma_i
\end{array}\right)(k)+ } \cr && \epsilon \left( \begin{array}{c}
F_i(k,\tv_i(k)) \\
\mu(\epsilon) \tanh\left(\frac{\tilde{U}_i(k)}{\tilde{S}_i(k)}\right) + \mu(\epsilon)\zeta_i(k)\\
\mu(\epsilon) \tilde{U}_i(k) \\
\mu(\epsilon) \tilde{S}_i(k)
\end{array} \right) + \epsilon \left(\begin{array}{c}
0 \\ z^{\vs}_i(k) \\ 0 \\ 0
\end{array}\right)
\end{eqnarray}
$i\in\mathcal{I}$, where $z^{\vs}_i(k)$ are correction terms for the operation-level updates that keeps them within the domain $[0,1]$. It is worth noting that the above recursion evolves in two timescales, the fast timescale of the operation-level update, $s_i(k)$ (including the approximations $\tilde{U}_i(k)$ and $\tilde{S}_i(k)$) and the slow timescale of the resource update, $\tv_i(k)$. For convenience, in several cases, we will denote $x_i(k)$ as the overall state vector of task $i$, i.e., $x_i(k)\df (\tv_i(k),s_i(k),\rho_i(k),\sigma_i(k))$ which evolves on $\mathcal{X}_i\df [0,1]\times[0,1]\times\mathbb{R}\times\mathbb{R}$. 

In the remainder of this paper, we will provide a characterization of the asymptotic behavior of the state profile $x(k)=(x_1(k),...,x_n(k))\in \mathcal{X}_1\times...\times\mathcal{X}_n$ as the time index $k$ increases. Note that the overall update recursion is stochastic in nature due to the presence of measurement noise $\eta_i(k)$ in the recordings of the performance of a task and secondly due to the artificial perturbation term, $\zeta_i(k)$, in the update of the operation-level. In the following analysis, we will often use the probability and expectation operator $\mathbb{P}_{x}$ and $\mathbb{E}_{x}$, initiated at state $x$, defined on the canonical path space generated by the sequences of the recursion (\ref{eq:OverallUpdateRecursion}) for each $i\in\mathcal{I}$.

\section{Resource Allocation Convergence Properties}
\label{sec:ResourceLevelConvergenceProperties}

In this section, we demonstrate the convergence properties of the resource update recursion (\ref{eq:RecursionForResources}) and \emph{independently of the operation-level update}. 

Before proceeding, it is important to derive bounds for the \emph{expected} utility measurement as well as the incremental difference of the resources. Let us introduce the notation: $\underline{\lambda}=\inf_{i\in\mathcal{I}}\lambda_i>0$, $\underline{c}=\inf_{i\in\mathcal{I}}c_i>1$, and $\overline{c} = \sup_{i\in\mathcal{I}}c_i>1$. We will also make frequent use of the following sets
$L_{\alpha} \df [0,\alpha)$ (i.e., `less than $\alpha$') and $G_{\alpha} \df (\alpha,1]$ (i.e., `greater than $\alpha$'), for some constant $\alpha\in(0,1)$.

\begin{proposition}[Bounded inverse utility]		\label{Pr:Property1}
As $\overline{\eta} \downarrow {0}$, 
\begin{equation*}
[\tilde{u}_i(k)]^{-1} \approx [u_i(s_i,v_i,d_i)]^{-1} + \ORDER{\overline{\eta}^2},
\end{equation*}
and
\begin{equation}	\label{eq:BoundsInverseUtility}
\frac{1}{\overline{c}} \leq [\tilde{u}_i(k)]^{-1} \leq 1 + \ORDER{\overline{\eta}^{2}},
\end{equation}
where $\ORDER{\cdot}$ denotes the order of the approximation error of the equality/inequality.
\end{proposition}
\ifproofs
\begin{proof}
By Taylor-series expansion of the inverse measurement function about its nominal value $u_i(s_i,v_i,d_i)$, we have that 
$$[\tilde{u}_i]^{-1} = \sum_{m\geq{0}}\frac{(-1)^{m}}{[u_i(s_i,v_i,d_i)]^{m+1}}\eta_i^{m}.$$ This approximation is convergent by the Ratio test (cf.,~\cite[Theorem~6.2.4]{Reed98}), since $\overline{\eta} < 1$ and $u_i(s_i,v_i,d_i) \geq {1}$. We conclude that 
\begin{eqnarray*}
[\tilde{u}_i]^{-1} & =  & \sum_{m\geq{0}}(-1)^{m}\frac{\eta_i^{m}}{[u_i(s_i,v_i,d_i)]^{m+1}} \cr
& \approx & [u_i(s_i,v_i,d_i)]^{-1} + \ORDER{\frac{\eta_i^2}{[u_i(s_i,v_i,d_i)]^3}}.
\end{eqnarray*}
The second part of the above approximation is nonnegative. Thus, given that $1 \leq u_i(s_i,v_i,d_i) \leq c_i \leq \overline{c}$, we conclude that 
\begin{equation*}
[\tilde{u_i}]^{-1} \geq 1/c_i \geq 1/\overline{c}.
\end{equation*}
Furthermore, given that $[u_i(s_i,v_i,d_i)]^{-1}$ is uniformly bounded from below, and the fact that $\eta_i^2\leq \overline{\eta}^2$, we may write equivalently that
\begin{eqnarray*}
[\tilde{u_i}]^{-1} & \approx & 
[u_i(s_i,v_i,d_i)]^{-1} + \ORDER{\eta_i^2}  \cr
& \leq & [u_i(s_i,v_i,d_i)]^{-1} + \ORDER{\overline{\eta}^2} \cr
& \leq & 1 + \ORDER{\overline{\eta}^2},
\end{eqnarray*}
which establishes the desired upper bound. 
\end{proof}
\fi

Define the quantities:
$\underline{\Lambda}(\tv_i) \df \underline{\lambda}/\overline{c} - v_i n (1 + \ORDER{\overline{\eta}^2})$ and 
$\overline{\Lambda}(\tv_i) \df - v_i n  \underline{\lambda}/\overline{c} + 1 + \ORDER{\overline{\eta}^2}.$

\begin{proposition}[Bounded fairness]	\label{Pr:Property2}
As $\overline{\eta}\downarrow{0}$, the incremental difference of the resource update satisfies
$\underline{\Lambda}(\tv_i) \leq F_{i}(k,\tv_i) \leq \overline{\Lambda}(\tv_i)$ for all $\tv_i\in[0,1]$.
\end{proposition}
\ifproofs
\begin{proof}
Given Proposition~\ref{Pr:Property1}, and as $\overline{\eta}\downarrow{0}$, we have
\begin{eqnarray*}
F_i(k,\tv_i) & = & \lambda_i [\tilde{u}_i(k)]^{-1} - \tv_i \sum_{j\in\mathcal{I}}\lambda_j[\tilde{u}_{j}(k)]^{-1} \cr
& \leq & \lambda_i\left(1+\ORDER{\overline{\eta}^2}\right) - v_i\sum_{j\in\mathcal{I}}\lambda_j/c_j \cr
& \leq & 1 + \ORDER{\overline{\eta}^2} - v_i n \underline{\lambda}/\overline{c},
\end{eqnarray*}
where the last inequality results from the fact that $\underline{\lambda}\leq\lambda_i\leq{1}$ and $\underline{c}\leq c_i \leq \overline{c}$. Accordingly, we get
\begin{equation*}
F_i(k,\tv_i) \geq \underline{\lambda}/\overline{c} - \tv_i n \left(1 + \ORDER{\overline{\eta}^2}\right),
\end{equation*}
which concludes the proof.
\end{proof}
\fi

\subsection{Feasibility}
\label{sec:Feasibility}

The first property of the proposed adjustment process is the \emph{feasibility} of the resulting vector of resources. In fact, we would like the resource vector $\vv(k)$ to remain within the probability simplex $\Simplex{n}$ for all future times $k$. 

\begin{proposition}[Feasibility] \label{Pr:Feasibility} Given a number of tasks $n\in\mathbb{N}$ and as $\overline{\eta}\downarrow{0}$, there exists $\epsilon^*=\epsilon^*(n,\overline{\eta})>0$, such that for any $\epsilon<\epsilon^*$, the update recursion (\ref{eq:RecursionForResources}) generates a sequence of resources $\{\vv(k)\}$ which satisfies $\vv(k)\in\Simplex{n}$ for all $k=1,2,...$ as long as $\vv(0)\in\Simplex{n}$.
\end{proposition}
\ifproofs
\begin{proof}
%
The sum of resources satisfies:
\begin{eqnarray*}
 \lefteqn{ \sum_{i=1}^{n}\tv_i(k+1)}\cr &=&  
 \sum_{i=1}^{n}v_i(k) + \epsilon \sum_{j=1}^{n}\lambda_j[\tilde{u}_j(k)]^{-1}\Big(1-\sum_{i=1}^{n}v_i(k)\Big)  .
\end{eqnarray*}
Note that the second part of the r.h.s. becomes identically zero when $\sum_{i=1}^{n}v_i(k)=1$. Thus, if the initial allocation satisfies $\sum_{i=1}^{n}v_i(0)=1$, then $\sum_{i=1}^{n}v_i(k)=1$ for all $k=1,2,...$. 

It remains to check under which conditions $\tv_i(k)\in[0,1]$. From Proposition~\ref{Pr:Property2}, we have that, for sufficiently small noise size $\overline{\eta}$, 
\begin{align*}
F_i(k,\tv_i) & \leq 1+\ORDER{\overline{\eta}^2} - \tv_i n \underline{\lambda}/\overline{c} \leq 1+\ORDER{\overline{\eta}^2} \cr
F_i(k,\tv_i) & \geq \underline{\lambda}/\overline{c} - \tv_i n (1+\ORDER{\overline{\eta}^2})  \geq  -n (1+\ORDER{\overline{\eta}^2}).
\end{align*}
Thus, the incremental difference of $\tv_i$ at time $k$ satisfies:
\begin{equation}
|\tv_i(k+1)-\tv_i(k)| \leq \epsilon n \left(1+\ORDER{\overline{\eta}^2}\right) \df \omega(\epsilon) > 0.
\end{equation}
As a result, and given the Markov property, in order for $\tv_i(k+1)$ to drop below zero, $\tv_i(k)$ should be at least within $\omega(\epsilon)$ distance from zero. Thus, it is sufficient to check the sign of the incremental difference of $\tv_i(k)$ when $\tv_i(k) \in [0,\omega(\epsilon))$. According to Proposition~\ref{Pr:Property2} and for any $\tv_i(k) \in [0,\omega(\epsilon))$, we have: 
\begin{equation*}
F_i(k,\tv_i(k)) \geq \underline{\lambda}/\overline{c} - \omega(\epsilon) n (1+\ORDER{\overline{\eta}^2}).
\end{equation*}
Given that $\underline{\lambda},\overline{c}>0$, there exists $\epsilon_1^*=\epsilon_1^*(n,\overline{\eta})$ sufficiently small, such that, for any $\epsilon < \epsilon_1^*$, we have that $F_i(k,\tv_i(k))\geq{0}$ uniformly on $\tv_i(k)\in[0,\omega(\epsilon))$.  In other words, the incremental difference points towards the interior of the domain.

Similarly, in order for $\tv_i(k+1)$ to become larger than $1$, $\tv_i(k)$ should be at least within $\omega(\epsilon)$-distance from $1$. Thus, it is sufficient to check the sign of the incremental difference of $\tv_i(k)$ when $\tv_i(k)\in(1-\omega(\epsilon),1]$. For any $\tv_i(k)\in(1-\omega(\epsilon),1]$, we have 
\begin{eqnarray*}
F_i(k,\tv_i(k)) & = & (1-\tv_i(k))\lambda_i [\tilde{u}_i(k)]^{-1} - \tv_i(k)\sum_{j\neq{i}}\lambda_j[\tilde{u}_j]^{-1} \cr 
& \leq & \omega(\epsilon)\lambda_i[\tilde{u}_i(k)]^{-1} - (1-\omega(\epsilon)) \sum_{j\neq{i}}\lambda_j[\tilde{u}_j]^{-1} \cr
& \leq & \omega(\epsilon)\left(1+\ORDER{\overline{\eta}^2}\right) - (1-\omega(\epsilon))\underline{\lambda}(n-1)/\overline{c}
\end{eqnarray*}
where we have used the properties $\underline{\lambda} \leq \lambda_i \leq{1}$ and $1/\overline{c} \leq [\tilde{u}_i]^{-1} \leq 1 + \ORDER{\overline{\eta}^2}$ for all $i\in\mathcal{I}$. Given that $\underline{\lambda},\overline{c}>0$, there exists $\epsilon^*_2=\epsilon_2^*(n,\overline{\eta})$ such that, for any $\epsilon<\epsilon_2^*(n,\overline{\eta})$, we have $F_i(k,\tv_i(k))\leq{0}$ for all $\tv_i(k)\in(1-\omega(\epsilon),1]$. In other words, the incremental difference points towards the interior of the domain.

In conclusion, for any $\epsilon<\epsilon^*\df\min\{\epsilon_1^*(n,\overline{\eta}),\epsilon_2^*(n,\overline{\eta})\}$, we have $\tv_i(k)\in[0,1]$ for any $k=1,2,...$ as long as $\tv(0)\in\Simplex{n}$.
\end{proof}
\fi

\textit{\textbf{For the remainder of the paper}}, we will assume that the step-size $\epsilon$ is chosen appropriately (according to Proposition~\ref{Pr:Feasibility}), so that the resource level is always within the feasible region for all tasks.

\subsection{Starvation Avoidance}
\label{sec:StarvationAvoidance}

The adjustment process guarantees \emph{starvation avoidance}, i.e., a positive amount of resources to all tasks and at all times. 

\begin{proposition}[Starvation Avoidance] \label{Pr:StarvationAvoidance}
Given a number of tasks $n\in\mathbb{N}$ and as $\overline{\eta}\downarrow{0}$, there exists $\alpha^*=\alpha^*(n)\df\underline{\lambda}/(n\overline{c})>0$ such that, for any task $i\in\mathcal{I}$, and any $0<\alpha\leq \alpha^*$, the following holds
$$\mathbb{P}_{x}\left[\liminf_{k\to\infty}{\rm dist}\left(\tv_i(k),G_{\alpha}\right) = 0 \right]=1.$$
\end{proposition}
\ifproofs
\begin{proof}
Let $\alpha>0$. We restrict the analysis to the per-task process $\{\tv_i(k)\}$. Let also consider the non-negative function $V(k,\tv_i) \df 1 - v_i \geq 0.$ The expected incremental difference of $V(k,\tv_i)$ satisfies 
\begin{eqnarray*}
\lefteqn{ \Delta V(k) } \cr 
& \df & \EXP{x}{V(k+1,\tv_i(k+1)) - V(k,\tv_i(k))|\vv_i(k)=\vv_i} \cr
& = & \EXP{x}{v_i(k) - v_i(k+1)|\tv_i(k)=\tv_i} \cr 
& = & - \epsilon \EXP{x}{F_i(k,v_i(k))|\tv_i(k)=\tv_i} \cr
& \leq & - \epsilon \underline{\Lambda}(\tv_i),
\end{eqnarray*}
for all $\tv_i\in[0,1]$, where $\underline{\Lambda}(\tv_i)$ is defined in Proposition~\ref{Pr:Property2}. For any $0<\delta<\alpha$ and any $\tv_i\in{L}_{\alpha-\delta}\df[0,\alpha-\delta)$, we have that: $$\underline{\Lambda}(\tv_i) \geq \underline{\lambda}/\overline{c} - (\alpha-\delta) n \left(1+\ORDER{\overline{\eta}^2}\right).$$ As $\overline{\eta}\downarrow{0}$, there exists $\alpha^*=\alpha^*(n)\df \underline{\lambda}/(n\overline{c})$ such that, if $\alpha \leq \alpha^*$, then we have
\begin{equation*}
\lim_{\overline{\eta}\downarrow{0}}\inf_{\tv_i\in{L}_{\alpha-\delta}}\underline{\Lambda}(\tv_i) = \underline{\lambda}/\overline{c} - (\alpha-\delta)n
\geq \delta n > 0
\end{equation*}
for any $\delta>0$. Then, the conclusion follows directly from \cite[Theorem~5.1]{Nevelson76}.
\end{proof}
\fi

Proposition~\ref{Pr:StarvationAvoidance} states that $\vv_i$ will approach infinitely often an amount of resources that it is at least $\alpha^*(n)>0$, i.e., it is always bounded away from zero for some fixed number of tasks $n$. Practically, this condition assures that zero amount of resources to one or more tasks cannot be sustainable.

\subsection{Balance}	\label{sec:Balance}

Another property of the resource update recursion (that is complementary to the \emph{starvation avoidance} property) assures that a task may never monopolize the available resources. This is  especially important in the case of a large number of tasks, thus establishing a form of \emph{balance} between tasks.

\begin{proposition}[Balance] \label{Pr:Balance}
For any number of tasks $n\in\mathbb{N}$ and as $\overline{\eta}\downarrow{0}$, there exists $\beta^*=\beta^*(n)=\overline{c}/(n\underline{\lambda})>0$, such that for any $\beta \geq \beta^*$,
$$\mathbb{P}_{x}\left[\liminf_{k\to\infty}{\rm dist}(\tv_i(k),L_{\beta}) = 0 \right]=1.$$
Note that as $n\to\infty$, then $\beta^*\downarrow{0}$ and $\beta^*n\to\overline{c}/\underline{\lambda}$.
\end{proposition}
\ifproofs
\begin{proof}
For any $\beta > 0$, let $\mathcal{I}'\subseteq{\mathcal{I}}$ be the set of tasks with resources greater than $\beta$, i.e., $\mathcal{I}'(\beta)\df \{i\in\mathcal{I}:v_i(k) > \beta\}$. For any $i\in\mathcal{I}'$, let us define the nonnegative function $V(k,v_i)\df v_i \geq 0$. The expected incremental difference of this function, as $\overline{\eta}\downarrow{0}$, satisfies 
\begin{eqnarray*}
\Delta{V}(k) & \df & \EXP{x}{v_i(k+1) - v_i(k)|\tv_i(k)=\tv_i} \cr & = &  \epsilon \EXP{x}{F_i(k,v_i(k))|\tv_i(k)=\tv_i} \cr
& \leq & \epsilon \overline{\Lambda}(\tv_i) \cr 
& = & \epsilon \left[ - \tv_i n \underline{\lambda}/\overline{c} + 1+\ORDER{\overline{\eta}^2} \right].
\end{eqnarray*}
For given $n$, there exists $\beta^*=\beta^*(n)=\overline{c}/(n\underline{\lambda})$ such that, for any $\beta\geq\beta^*$, we have
$$\lim_{\overline{\eta}\downarrow{0}}\sup_{\tv_i(k)\in G_{\beta+\delta}}\overline{\Lambda}(\tv_i) \leq 1 - \frac{\beta+\delta}{\beta^*} < 0,$$ for any $\delta>0$. According to \cite[Theorem~5.1]{Nevelson76}, the conclusion follows. Finally, note that as $n\to\infty$, then $\beta^*(n)\downarrow{0}$ and $\beta^* n \to \overline{c}/\underline{\lambda}$.
\end{proof}
\fi

Proposition~\ref{Pr:Balance} states that, for any number of tasks $n$, there exists $\beta^*=\beta^*(n)>0$, such that, for any $\beta\geq\beta^*$ and independently of the initial conditions, the process will visit the set $L_\beta=[0,\beta)$ infinitely often, i.e., the resources of each task will drop below $\beta$. Furthermore, note that $\beta^*$ approaches zero as the number of tasks increases. Informally, we may say that \emph{no task can monopolize the available resources when the number of tasks increases}.

\subsection{Fairness}     \label{sec:Fairness}

Lastly, we demonstrate one of the most attractive properties of the proposed resource update recursion, i.e., the fact that, for any given operation-level profile $s$ and demand profile $d$, the allocation of resources will approach a \emph{fair allocation}, as defined by Definition~\ref{def:EfficientAllocation}. This is formally stated as follows. For some fixed operation-level and demand profiles, $s\in\mathcal{S}$ and $d\in\mathcal{D}$, define the set:
\begin{equation*}
\mathcal{F} = \mathcal{F}(s,d) \df \left\{\vv\in\cV: \Phi_{i}(\tv,s,d) = 0\,, \forall i\in\mathcal{I} \right\},
\end{equation*}
which contains the fair allocations, according to Definition~\ref{def:EfficientAllocation}. 
\begin{proposition}[Fairness]	\label{Pr:Fairness}
For any fixed operation-level profile $s\in\mathcal{S}$ and demand profile $d\in\mathcal{D}$, and as $\overline{\eta}\downarrow{0}$ and $\epsilon\downarrow{0}$ $$\Prob{x}{\liminf_{k\to\infty}{\rm dist}\left(\vv(k),\mathcal{F}\right)=0}=1.$$
\end{proposition}
\ifproofs
\begin{proof}
Let us define the nonnegative function
\begin{equation*}
W(k,\vv) \df \sum_{i\in\mathcal{I}}F_i(k,\tv_i)^2 \geq 0.
\end{equation*}
For each $k$, $\tilde{u}_i(k)$ constitutes exogenous factors of the function $F_i$, which is continuously differentiable with respect to $\tv_i$. We can approximate the expected incremental gain of $W(k,\vv)$ by applying a Taylor series expansion. All expectations in this proof are conditioned to $\vv(k)=\tv$. We have:
\begin{eqnarray*}
\lefteqn{\Delta{W}(k) }\cr 
& \df & \mathbb{E}_{x} \Big[ \sum_{i\in\mathcal{I}}\left[F_i(k+1,\tv_{i}(k+1))^2 - F_i(k,\tv_i(k))^2 \right] \Big] \cr 
& \approx & \epsilon \sum_{i\in\mathcal{I}}\mathbb{E}_{x}\Big[[\nabla_{v_i} F_i(k,\tv_i)^2]\tr F_i(k,\tv_i)\Big]
\end{eqnarray*}
plus higher order terms of $\epsilon$ and $F_i(k,\tv_i)$. Note that $$\nabla_{\tv_i}[F_{i}(k,\tv_i)^2]=-2F_i(k,\tv_i)\sum_{j\in\mathcal{I}}\lambda_j[\tilde{u}_{j}(k)]^{-1}.$$
Thus,
\begin{eqnarray*}
\lefteqn{\Delta{W}(k) }\cr
& \approx & \epsilon \sum_{i\in\mathcal{I}} \mathbb{E}_{x} \Big[\Big(-2\sum_{j\in\mathcal{I}}\lambda_j[\tilde{u}_j(k)]^{-1}\Big)F_i(k,\tv_i)^2 \Big] \cr 
& = & -2\epsilon\sum_{i\in\mathcal{I}} \sum_{j\in\mathcal{I}}\lambda_j\EXP{x}{[\tilde{u}_{j}(k)]^{-1}F_{i}(k,\tv_i)^{2}}.
\end{eqnarray*}
Given the boundedness of the performance function and the measurement noise (\ref{eq:NoisyObservations}), we have that $\sup_{\tv_i\in[0,1]}\tilde{u}_{i} = c_i + \overline{\eta}$, which results in $$\inf_{\tv_i\in[0,1]}[\tilde{u}_i]^{-1} = \frac{1}{c_i+\overline{\eta}} \geq \frac{1}{\overline{c}+\overline{\eta}}.$$ Thus, we have that
\begin{equation*}
\Delta{W}(k) \leq \frac{-2\epsilon}{\overline{c}+\overline{\eta}}\sum_{i\in\mathcal{I}} \sum_{j\in\mathcal{I}}\lambda_j\EXP{x}{F_{i}(k,\tv_i)^{2}}.
\end{equation*}
Let us define the set 
$$\tilde{\mathcal{F}} \df \left\{\vv\in\Simplex{n}: \EXP{x}{F_{i}(k,\tv_i)^2} = 0, \forall i\in\mathcal{I}\right\}.$$ For some $\delta>0$, let $\mathcal{G}_{\delta} \df \Simplex{n}\backslash\mathcal{B}_{\delta}(\tilde{\mathcal{F}})$, i.e., the set $\mathcal{G}_{\delta}$ contains all non-efficient allocations that are at least $\delta$ far from the ones in $\tilde{\mathcal{F}}$. It is evident that $$\inf_{\vv\in \mathcal{G}_{\delta}}\sum_{i\in\mathcal{I}}\EXP{x}{F_i(k,\tv_i)^2}>0\,, \mbox{ for all } \delta>0.$$ Thus, for sufficiently small $\epsilon>0$, $\sup_{\vv\in\mathcal{G}_{\delta}}\Delta{W}(k)<0$ for all $\delta>0$. By \cite[Theorem~5.1]{Nevelson76}, we have that, for sufficiently small $\epsilon>0$, $$\Prob{x}{\liminf_{k\to\infty}{\rm dist}\left(\vv(k),\tilde{\mathcal{F}}\right)=0}=1.$$ 
By convexity of the function $x^2$ and Jensen's inequality, we also have that 
\begin{eqnarray*}
\EXP{x}{F_i(k,\tv_i(k))^2} & \geq & \left(\EXP{x}{F_i(k,\tv_i(k))}\right)^2 \cr
& \approx & \left( \Phi_i(\tv,s,d) + \ORDER{\overline{\eta}^2} \right)^2,
\end{eqnarray*}
for all $i\in\mathcal{I}$. Thus, as $\overline{\eta}\downarrow{0}$, $\tilde{\mathcal{F}} \subseteq \mathcal{F}$, which concludes the proof.
\end{proof}
\fi

Proposition~\ref{Pr:Fairness} states that, for any fixed operation-level and demand profile, the allocation $\vv$ of resources will reach the set of fair allocations $\mathcal{F}$ infinitely often with probability one. Furthermore, this fairness guarantee holds independently of the number of tasks and the specifics of their utility functions. 

\subsection{Discussion}

The above properties of \emph{feasibility}, \emph{starvation avoidance}, \emph{balance} and \emph{fairness} of Propositions~\ref{Pr:Feasibility}, \ref{Pr:StarvationAvoidance}, \ref{Pr:Balance} and \ref{Pr:Fairness}, respectively, provide guarantees that any scheduling mechanism should provide independently of the application of interest. For example, in CPU Bandwidth Allocation example of Section~\ref{sec:ExampleCPUBandwidthAllocation}, these properties are essential and should be satisfied by any operating system, e.g., tasks should always receive resources and resources should be balanced. It is important to also note that these properties were shown only under Assumption~\ref{As:UtilityFunction}.

\section{Overall Convergence Properties}		\label{sec:OverallConvergenceProperties}

The previous convergence results were derived without imposing any specific conditions on the operation-level $s_i(k)$, $i\in\mathcal{I}$. In fact, the conclusions of Propositions~\ref{Pr:Feasibility}--\ref{Pr:Balance} were independent of the operation-level profile, while Proposition~\ref{Pr:Fairness} assumes a fixed operation-level profile. In this section, we wish to provide a more detailed characterization of the global attractors when the operation-levels are also adjusted according to (\ref{eq:OverallUpdateRecursion}). 

\subsection{ODE approximation}

The asymptotic behavior of the overall recursion (\ref{eq:OverallUpdateRecursion}) can be analyzed by the ODE-method for stochastic approximations \cite{KushnerYin03}. In particular, the convergence behavior can be associated with the limit points of the following system of ordinary differential equations (ODE's):
\begin{equation} \label{eq:OverallODE}
 \left(\begin{array}{c}
    \dot{\bar{v}}_i \\ \dot{\bar{s}}_i \\ \dot{\bar{\rho}}_i \\ \dot{\bar{\sigma}}_i
  \end{array}
    \right) = \left(\begin{array}{c} \Phi_i(\bar{s},\bar{\vv},\bar{\vd}) \\ 
    \mu \tanh\left(\frac{u_i(\bar{s}_i,\bar{\tv}_i,\bar{d}_i) - \rho_i}{\bar{s}_i - \bar{\sigma}_i}\right) + Z_i^{s} \\
    \mu \gamma \left(u_i(\bar{s}_i,\bar{\tv}_i,\bar{d}_i) - \bar{\rho}_i\right) \\
    \mu \gamma \left(\bar{s}_i - \bar{\sigma}_i\right) \end{array}\right),
\end{equation}
for each $i\in\mathcal{I}$, as the step-size $\epsilon$ approaches zero. The paths $\bar{\vv}_i(t)$, $\bar{s}_i(t)$, $\bar{d}_i(t)$, $\bar{\rho}_i(t)$, and $\bar{\sigma}_i(t)$ will be associated with the weak-limits of the linear-time interpolations\footnote{The linear-time interpolation of a sequence $\vv(k)$, $k=0,1,...$, is defined as $\overline{\vv}^{\epsilon}(\tau)=\vv(k)$ for all $\epsilon t_k\leq{\tau}<\epsilon t_{k+1}$.} of the discrete-time sequences $\tv_i(k)$, $s_i(k)$, $d_i(k)$, $\rho_i(k)$ and $\sigma_i(k)$, respectively. The scalar $Z_i^s$ corresponds to the minimum effort required to drive $\bar{s}_i(t)$ back to $[0,1]$. All terms of the above ODE are functions of an artificial continuous-time index $t$, generated by the interpolated $\epsilon$-dependent time-scale (cf.,~\cite[Chapter~8]{KushnerYin03}).

\subsection{Global convergence}	\label{sec:GlobalConvergence}

Before proceeding on establishing the characterization of the limiting properties of the original discrete-time process (\ref{eq:OverallUpdateRecursion}) with the ODE approximation (\ref{eq:OverallODE}), the following lemma provides a characterization of the solutions of the ODE as $\gamma$ increases.

\begin{lemma}[Continuous-function approximation]	\label{Lm:Approximators}
Let $(\bar{v}_i^{\gamma},\bar{s}_i^{\gamma},\bar{\rho}_i^{\gamma},\bar{\sigma}_i^{\gamma})$ denote the $\gamma$-dependent solutions to (\ref{eq:OverallODE}). Let also $\bar{u}_i^{\gamma}(t)\df u_i(\bar{v}_i^{\gamma},\bar{s}_i^{\gamma},\bar{d}_i)$ denote the performance function evaluated along the $\gamma$-dependent solutions. For any compact interval $[t_1,t_2]\subset\mathbb{R}_{+}$, with $t_1>0$, there exist an unbounded increasing sequence $\{\gamma_k\}$ and absolutely continuous functions $\varrho_i$ and $\varsigma_i$ with derivatives $\dot{\varrho}_i$ and $\dot{\varsigma}_i$, respectively, such that
\begin{enumerate}
\item $\bar{u}_i^{\gamma_k}(t)$ and $\bar{\rho}_i^{\gamma_k}$ converge to $\varrho_i$ uniformly;
\item $\dot{\bar{u}}_i$ and $\dot{\bar{\rho}}_i^{\gamma_k}$ converge weakly to $\dot{\varrho}_i$ in $L^1([t_1,t_2],\mathbb{R})$.
\item $\bar{s}_i^{\gamma_k}$ and $\bar{\sigma}_i^{\gamma_k}$ converge to $\varsigma_i$ uniformly;
\item $\dot{\bar{s}}_i^{\gamma_k}$ and $\dot{\bar{\sigma}}_i^{\gamma_k}$ converge weakly to $\dot{\varsigma}_i$ in $L^1([t_1,t_2],\mathbb{R})$.
\end{enumerate}
\end{lemma}
\ifproofs
\begin{proof}
Note that both $\bar{u}_i^{\gamma}(t)$ and $\bar{s}^{\gamma}(t)$ are uniformly bounded. Given that the corresponding derivatives $\dot{\bar{u}}_i^{\gamma}(t)$ and $\dot{\bar{s}}^{\gamma}(t)$ are formed using bounded elements, they are also uniformly bounded. According to the ODE~(\ref{eq:OverallODE}), we have:
\begin{equation*}
\dot{\bar{u}}_i - \dot{\bar{\rho}}_i = \dot{\bar{u}}_i - \mu\gamma (\bar{u}_i - \bar{\rho}_i).
\end{equation*}
Using standard Lyapunov stability analysis (cf.,~\cite[Section~4.9]{Khalil92}), the following approximation holds:
\begin{equation*}
\magn{\bar{u}_i^{\gamma}(t)-\bar{\rho}_i^{\gamma}(t)} \leq e^{-\mu\gamma{t}}\magn{\bar{u}_i^{\gamma}(0)-\bar{\rho}_i^{\gamma}(0)} + \frac{1}{\mu\gamma}\sup_{\tau>0}\magn{\dot{\bar{u}}_i^{\gamma}(\tau)}
\end{equation*}
which implies that $\bar{\rho}_i^{\gamma}$ and $\dot{\bar{\rho}}_i^{\gamma}$ are also uniformly bounded over any bounded interval $[t_1,t_2]\subset\mathbb{R}_+$. Thus, for any increasing unbounded sequence $\gamma_k$, the set of functions $\{\bar{u}_i^{\gamma_k}-\bar{\rho}_i^{\gamma_k}\}_k$ is \emph{equicontinuous in the extended sense} (cf.,~\cite[Section~4.2]{KushnerYin03}). Then, according to \cite[Theorem~4]{Aubin84}, there exists a subsequence (which we relabel $\gamma_k$) such that, for any bounded interval $[t_1,t_2]$, a) the sequence of functions $\{\bar{u}_i^{\gamma_k}-\bar{\rho}_i^{\gamma_k}\}_k$ converges uniformly on a absolutely continuous function $\Delta{\varrho}_i(t)$, and b) the sequence of functions $\{\dot{\bar{u}}_i^{\gamma_k}-\dot{\bar{\rho}}_i^{\gamma_k}\}_k$ converges weakly to $\dot{\overline{\Delta{\varrho}_i}}(t)$ in $L_1([t_1,t_2],\mathbb{R})$. Thus, the sequence
\begin{equation*}
\frac{1}{\mu\gamma_k}\dot{\bar{\rho}}_i^{\gamma_k}(t) = ( \bar{u}_i^{\gamma_k} - \bar{\rho}_i^{\gamma_k} )
\end{equation*}
converges uniformly to $\Delta{\varrho}_i(t)$. Since, $\gamma_k$ is unbounded, and $\dot{\bar{\rho}}_i^{\gamma_k}$ is uniformly bounded, we conclude that $\Delta{\varrho}_i(t)\equiv{0}$. This further implies that $\dot{\overline{\Delta{\varrho}_i}}\equiv{0}$. Thus, the conclusions of (1) and (2) immediately follow.
Following similar reasoning, the conclusions (3) and (4) can also be derived.
\end{proof}
\fi

Using Lemma~\ref{Lm:Approximators}, we can derive a detailed characterization of the potential attractors of the discrete-time recursions (\ref{eq:OverallUpdateRecursion}), as the following theorem demonstrates.

\begin{theorem}[Global convergence]	\label{Th:GlobalConvergence}
Consider the overall update recursion (\ref{eq:OverallUpdateRecursion}) with a fixed demand $\vd(k)=\vd$ and a step-size $\epsilon$ satisfying condition (\ref{eq:StepSizeConstraints}) and $0 < \epsilon\mu(\epsilon) < \nicefrac{1}{\gamma}.$ Let $\mathcal{L}$ denote the limit points\footnote{The set of \emph{limit points} $\mathcal{L}$ of an ODE $\dot{x}=g(x)$ with domain $A$ is defined as $\mathcal{L}\df \lim_{t\to\infty}\bigcup_{x\in{A}}\{x(s),s\geq{t}:x(0)=x\}$, i.e., it is the set of all points in $A$ to which the solution of the ODE converges.} of the system of ODE's
\begin{equation}	\label{eq:LimitingODE}
\dot{\bar{\tv}}_i(t) = \Phi_i\left(\bar{s}^*(\bar{\vv},\bar{d}),\bar{\vv},\bar{d}\right), \quad i\in\mathcal{I},
\end{equation} 
where $\bar{s}^*=(\bar{s}^*_1,...,\bar{s}^*_n)$, and 
\begin{equation} \label{eq:OperationLevelMaximizers}
\bar{s}_i^*(\bar{\tv}_i,\bar{d}_i) \df \arg\max_{\bar{s}_i\in[0,1]}u_i(\bar{s}_i,\bar{\tv}_i,\bar{d}_i), \quad i\in\mathcal{I}.
\end{equation}
As $\overline{\eta}\downarrow{0}$ and $\gamma\to\infty$, the following hold: for any $\delta>0$, the fraction of time that the discrete-time process $\{s(k),\vv(k)\}$ spends in the $\delta$-neighborhood of $\mathcal{L}$, $\mathcal{B}_{\delta}(\mathcal{L})$, goes to one (in probability) as $k\to\infty$.
\end{theorem}
\ifproofs
\begin{proof}
Note the following:
\begin{itemize}
\item The utility function $u_i(\bar{s}_i,\bar{v}_i,\bar{d}_i)$ is continuous with respect to the operation-level $s_i$ and the resource level $\tv_i$, and therefore $\Phi(\cdot,\cdot,\bar{d})$ is also continuous with respect to $s$ and $\vv$. 
\item The observation terms in (\ref{eq:OverallUpdateRecursion}) are uniformly bounded for all $k$ in the domain, and therefore uniformly integrable. To show this, first note that $\EXP{x}{\magn{F_i(k,\tv_i(k))}}<\infty$ uniformly for all $k$, according to Proposition~\ref{Pr:Property2}. Secondly, note that $$\magn{\rho_i(k)-\rho_i(0)} \leq \sum_{\ell=0}^{k}\kappa(1-\kappa)^{k}\magn{\tilde{u}_i(k-\ell)-\rho_i(0)},$$ where $\kappa\df\epsilon\mu(\epsilon)\gamma$. Given Proposition~\ref{Pr:Property1}, if $\kappa < 1$ (which will be the case when we take $\epsilon\downarrow{0}$), $\magn{\rho_i(k)-\rho_i(0)}<\infty$ uniformly for all $k$ as long as $\rho_i(0)$ is bounded. The same conclusion can also be derived for the auxiliary state variable $\sigma_i(k)$, given that $s_i(k)\in[0,1]$ for all $k$.
\item Let $\mathcal{L}'$ denote the limit points of the ODE~(\ref{eq:OverallODE}). The uniform integrability of the observation terms in (\ref{eq:OverallUpdateRecursion}) establishes (according to Theorem 8.2.1 in \cite{KushnerYin03}) the following weak-convergence: As $\overline{\eta}\downarrow{0}$ and for any $\delta>0$, the fraction of time that $(s_i(k),\tv_i(k),\rho_i(k),\sigma_i(k))_i$ spends in the $\delta$-neighborhood of $\mathcal{L}'$, $\mathcal{B}_{\delta}(\mathcal{L}')$, goes to one (in probability) as $\epsilon\downarrow{0}$ and $k\to\infty$.

\item Consider the (unprojected) faster response ODE, defined by
\begin{equation} \label{eq:FasterScaleODE}
 \left(\begin{array}{c}
    \dot{\bar{s}}_i \\ \dot{\bar{\rho}}_i \\ \dot{\bar{\sigma}}_i
  \end{array}
    \right) = \left(\begin{array}{c}  
    \mu \tanh\left(\frac{u_i(\bar{s}_i,\bar{\vv}_i,\bar{d}_i) - \bar{\rho}_i}{\bar{s}_i - \bar{\sigma}_i}\right) \\
    \mu \gamma \left(u_i(\bar{s}_i,\bar{\tv}_i,\bar{d}_i) - \bar{\rho}_i\right) \\
    \mu \gamma \left(\bar{s}_i - \bar{\sigma}_i\right) \end{array}\right),
\end{equation}
for some given $\bar{\vv}$ and $\bar{d}$. Furthermore, consider the nonnegative function
$$W(\bar{s}) \df \sum_{i\in\mathcal{I}}\left\{ \max_{s_i\in[0,1]}u_i(s_i,\bar{\tv}_i,\bar{d}_i) - u_{i}(\bar{s}_i,\bar{\tv}_i,\bar{d}_i)\right\}.$$ Let us denote by $(\bar{s}_i^{\gamma},\bar{\rho}_i^{\gamma},\bar{\sigma}_i^{\gamma})$ to be the $\gamma$-dependent solution of the faster-response ODE~(\ref{eq:FasterScaleODE}). Its time derivative, calculated along the solution, satisfies:
\begin{equation*}
\dot{W}(\bar{s}^{\gamma}) = - \mu \sum_{i\in\mathcal{I}}\nabla_{\bar{s}_i^{\gamma}}u_i(\bar{s}_i^{\gamma},\bar{\tv}_i,\bar{d}_i) \cdot \dot{\bar{s}}^{\gamma}_i(t).
\end{equation*}
Given Lemma~\ref{Lm:Approximators}, the solution of the ODE satisfies: 
\begin{equation*}
\dot{\bar{s}}_i^{\gamma}(t) = \mu\tanh\left(\frac{\dot{\bar{\rho}}^{\gamma}_i(t)}{\dot{\bar{\sigma}}^{\gamma}_i(t)}\right) \xRightarrow {\gamma\to\infty}{} \mu\tanh\left( \frac{\dot{\varrho}_i(t)}{\dot{\varsigma}_i(t)} \right),
\end{equation*}
where convergence is in the weak sense with respect to the norm $L^1([t_1,t_2],\mathbb{R})$, for every bounded interval $[t_1,t_2]$. Furthermore, $\bar{u}_i^{\gamma}(t) \df u_i(\bar{s}^{\gamma}_i(t),\bar{v}_i,\bar{d}_i)$ converges uniformly to $\varrho_i(t) = u_i(\varsigma_i(t),\bar{v}_i,\bar{d}_i)$ as $\gamma$ increases. Since $u_i$ is continuously differentiable with respect to the operation-level, we also have (using the chain-rule)
\begin{equation*}
\dot{\varrho}_i(t) = \left(\nabla_{\varsigma_i}u_i(\varsigma_i,\bar{v}_i,\bar{d}_i)\right) \dot{\varsigma}_i(t).
\end{equation*}
Thus, we conclude that, 
\begin{eqnarray*}
\lefteqn{\dot{W}(\bar{s}^{\gamma}) \xRightarrow{\gamma\to\infty} } \cr && -\mu\sum_{i\in\mathcal{I}} \left(\nabla_{\varsigma_i}u_i(\varsigma_i,\bar{v}_i,\bar{d}_i)\right) \tanh\left(\nabla_{\varsigma_i}u_i(\varsigma_i,\bar{v}_i,\bar{d}_i)\right) \leq {0}.
\end{eqnarray*}
The time derivative of the nonnegative function $W(\cdot)$ accepts a unique zero, satisfying condition $\nabla_{\varsigma_i}u_i(\varsigma_i,\bar{v}_i,\bar{d}_i)=0$. According to \cite[Theorem~3.1]{Khalil92}, the operation-level satisfying condition (\ref{eq:OperationLevelMaximizers}) is globally asymptotically stable equilibrium point of the fast response dynamics (\ref{eq:FasterScaleODE}). Given also Assumption~\ref{As:UtilityFunction}, $\bar{s}_i^*(\bar{\tv}_i,\bar{d}_i)$ is the unique globally asymptotically stable equilibrium point of the ODE~(\ref{eq:FasterScaleODE}).
\item Given that the globally asymptotically stable equilibrium of the unprojected dynamics lies within the domain $\mathcal{S}_i\df[0,1]$, $\bar{s}_i^*$ is also the unique globally asymptotically stable point of the projected ODE.
\end{itemize}

Thus, the conclusion is a direct implication of \cite[Theorem~8.6.1]{KushnerYin03}.
\end{proof}
\fi

\subsection{Discussion} \label{sec:DiscussionGlobalConvergence}

The importance of Theorem~\ref{Th:GlobalConvergence} is two-fold. First, we guarantee that independently of the resource adjustments $\tv_i(k)$, the operation-level is always located at a maximizer of the utility function $u_i$ (i.e., at an \emph{efficient} allocation in $\mathcal{E}^*$). Second, the allocation of resources converges \emph{in distribution} to the limit points of ODE~(\ref{eq:LimitingODE}), i.e., the fraction of time that this occurs approaches one as the step-size approaches zero and the number of iterations increases. The limit points of this ODE contain all \emph{fair} allocations of $\mathcal{F}^*$. 

In general, the set of limit points of the limiting ODE~(\ref{eq:LimitingODE}) may not necessarily be restricted to the set of fair allocations $\mathcal{F}^*$. In order to provide a more detailed characterization of the limit points, additional information over the utility functions $u_i$ should be available. However, independently of the availability of such additional information, Propositions~\ref{Pr:Feasibility}, \ref{Pr:StarvationAvoidance} and \ref{Pr:Balance} provide generic bounds for the set of limit points. For example, according to Proposition~\ref{Pr:Balance}, the discrete time process visits infinitely often the set $L_{\beta^*}$ (balanced allocations), which (by Theorem~\ref{Th:GlobalConvergence}) further implies that the set of limit points should also be contained within the set $L_{\beta^*}$.

Further note that Theorem~\ref{Th:GlobalConvergence} applies for a fixed demand $d=(d_1,...,d_n)$ set by the user. However, if the demand changes, the algorithm will automatically adapt to the new conditions, due to the use of a constant step-size $\epsilon$. That is, the algorithm is adaptive to changes in the demand of the user. The time needed for the algorithm to converge to the efficient allocations depend on the selected step-size $\epsilon$. 

Given the linear complexity of the recursions (\ref{eq:OverallUpdateRecursion}) with the number of tasks, the algorithm is computationally efficient. The recursion (\ref{eq:OverallUpdateRecursion}) may also be implemented independently by each task $i$, thus distributing the computational burden  from the \RM\ to the tasks. In this case, the \RM\ is simply responsible for communicating the measured quantities $F_i$ to each one of the tasks.

\section{Simulations}	\label{sec:Simulations}

In this section, we provide a simulation study to demonstrate the convergence properties of the proposed dynamics. In all considered simulation studies, we introduce 3 time-zones (depicted by (e1), (e2) and (e3)), where we alter the demand requested by the tasks. In particular, in time-zone (e2) the demand of half the considered tasks increases by a factor of 2, while in time-zone (e3) the demand of the same tasks returns to its initial level (i.e., the one at time-zone (e1)). With this variation in the demand of the tasks, we wish to demonstrate the adaptation of the proposed learning framework in varying user requests.

\begin{figure}[th!]	
\iffigures
\centering
\includegraphics[scale=0.63]{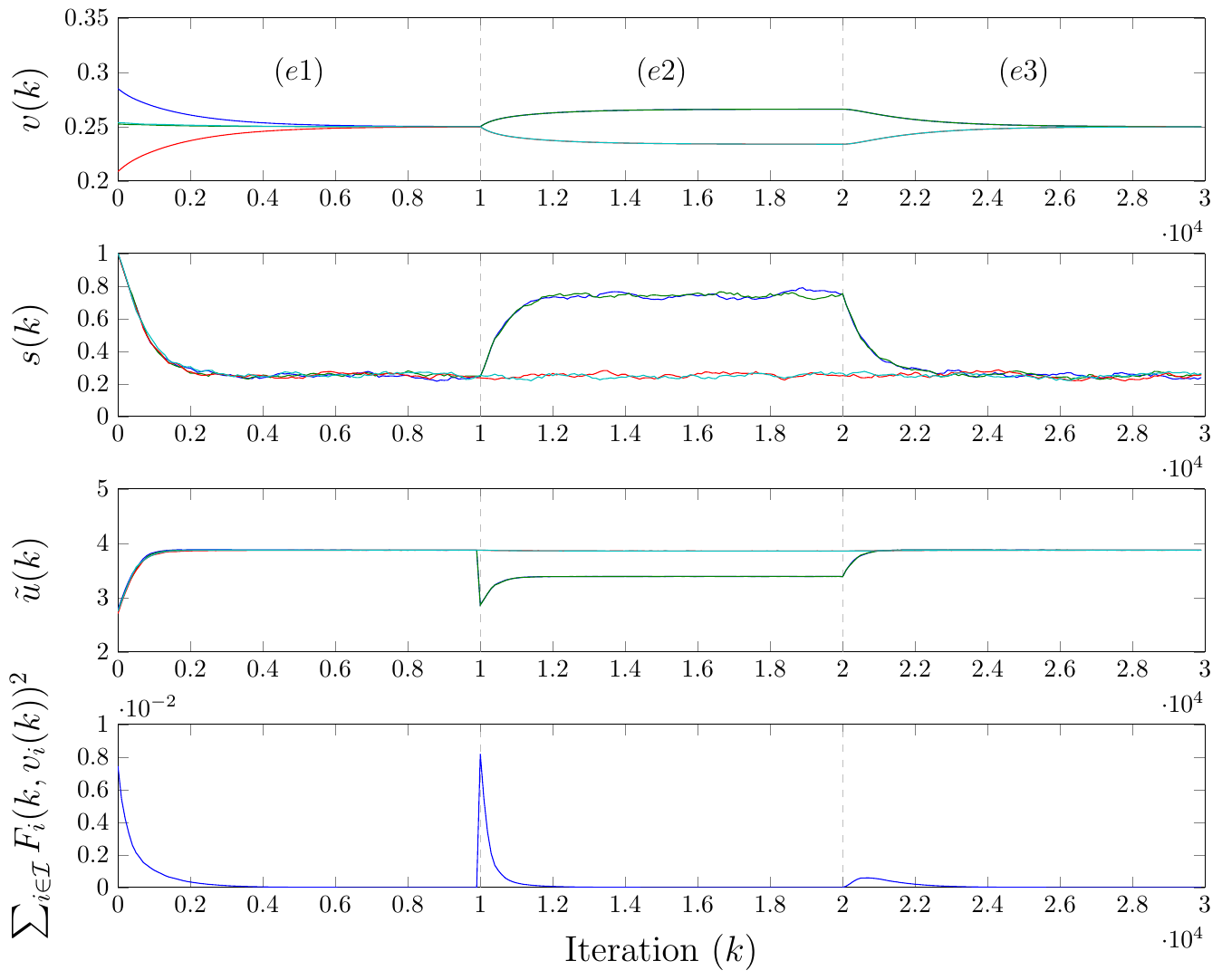}
\caption{Resource and operation-level adaptation for 4 identical tasks.}
\label{fig:3identicaltasks}
\fi
\end{figure}

\emph{In the first simulation study of Figure~\ref{fig:3identicaltasks}}, we consider 4 \emph{identical} tasks with a utility function of the form~(\ref{eq:UtilityHomeEnergyManagement}), i.e., all parameters of the utility function are identical in each task. Furthermore, the initially requested demands are identical, as well as the weights of the tasks, i.e., $\lambda_i=\lambda=1$ for each $i$. In particular, the considered parameters of the tasks simulated are: $a_i=2$, $b_i=1$, $c_i=2$, $\epsilon=0.0005$, $\mu(\epsilon)=\epsilon^{-1/20}$, and $\overline{\eta}=\overline{\zeta}=0.001$. In this case, and according to Definition~\ref{def:EfficientAllocation}, we should expect a unique efficient allocation corresponding to $\lambda_i/\sum_{j}\lambda_j=1/n=1/4=0.25$. This is indeed the emergent behavior in time-zone (e1). When the demand increases for two of the tasks, the emergent allocation accommodates this request (time-zone (e2)), and when the updated requests return to their original values, the initial allocation emerges again. This adaptive response of the dynamics to the demand variations should be attributed to the selection of constant step-size $\epsilon$, and agrees with our remark in Section~\ref{sec:DiscussionGlobalConvergence}. Note, finally, that throughout the simulation study, the efficiency criterion of $F_i$ approaching zero is maintained, something that verifies our global convergence result of Theorem~\ref{Th:GlobalConvergence}.

\begin{figure}[!ht]	
\iffigures
\centering
\includegraphics[scale=0.63]{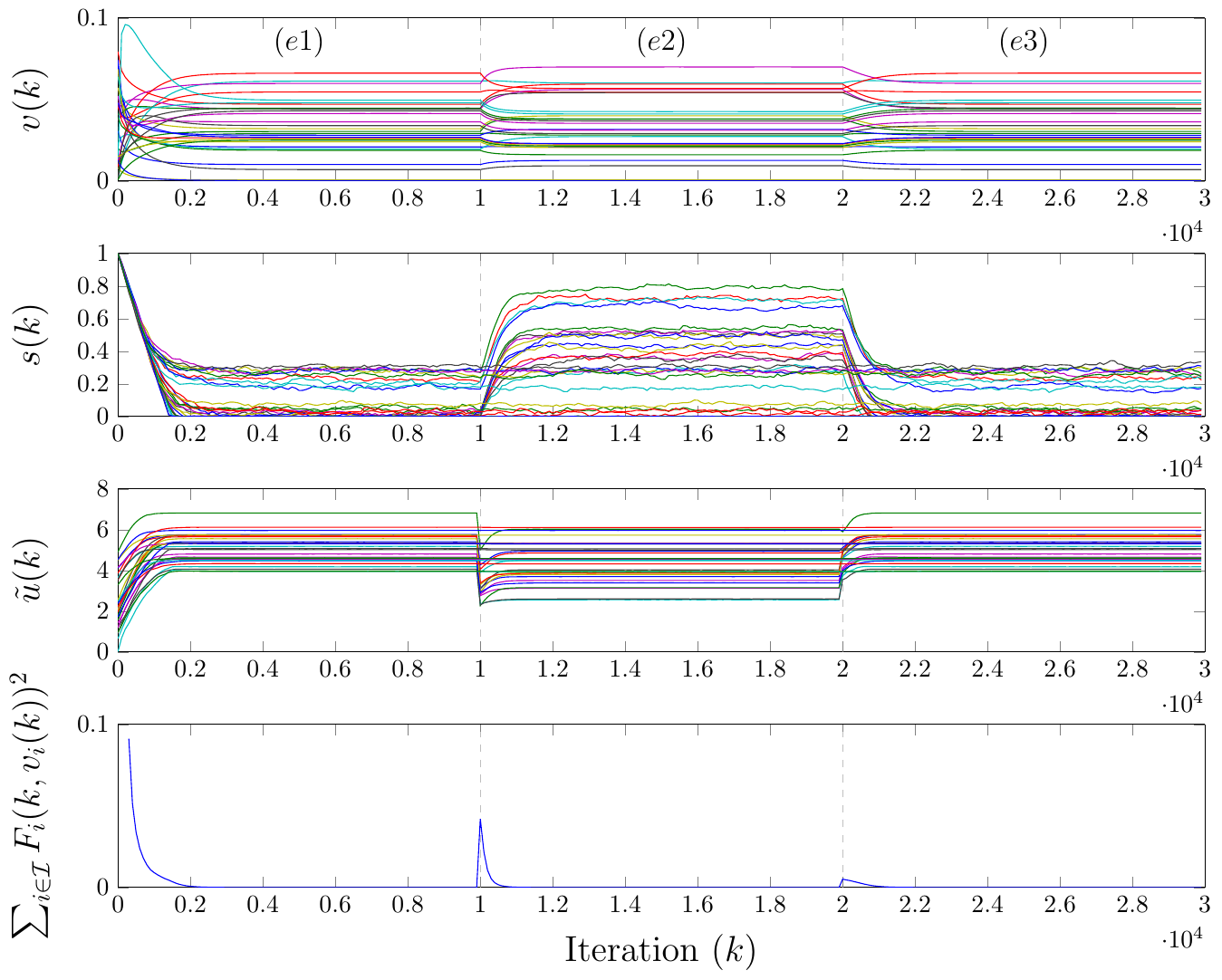}
\caption{Resource and operation-level adaptation for 30 randomly generated tasks.}
\label{fig:30tasks}
\fi
\end{figure}

\emph{In the second simulation study of Figure~\ref{fig:30tasks}}, we consider 30 tasks with a utility function of the form~(\ref{eq:UtilityHomeEnergyManagement}). The parameters of the utility function, the weights and the demands of the tasks are randomly generated. Note that the main conclusions noted for the simple case of 4 tasks continue to hold even for this large number of tasks. In particular, the dynamics adapt rather fast to the variation in the demands, while the efficiency criterion is maintained throughout the simulations.

\section{Conclusions and Future Work} \label{sec:Conclusions}

We proposed a measurement- or performance-based learning scheme for addressing a large class of resource allocation problems. An initially formulated centralized objective was translated into resource and operation-level adjustment dynamics which exhibits desirable properties, such as feasibility, starvation avoidance, balance and fairness/efficiency. Furthermore, global convergence guarantees to fair/efficient outcomes were demonstrated under generic assumptions in the design of the utility functions. 

The importance of the proposed methodology lies in the fact that no a-priori knowledge of the details of the utility function is required, something that it is relevant to several practical scenarios, such as the allocation of CPU bandwidth in computing applications. The proposed dynamics can also be distributed in a natural way, since implementation can be performed by the tasks rather than the \RM, a property that is rather attractive when we consider an extremely large number of applications. 

\bibliographystyle{plain} 
\bibliography{Bibliography}

\begin{thebibliography}{10}

\bibitem{DeAngelis13}
F.~De Angelis, M.~Boaro, D.~Fuselli, S.~Squartini, F.~Piazza, and Q.~Wei.
\newblock Optimal home energy management under dynamic electrical and thermal
  constraints.
\newblock {\em IEEE Transactions on Industrial Informatics}, 9(3):1518--1527,
  Aug 2013.

\bibitem{Aubin84}
J.~P. Aubin and A.~Cellina.
\newblock {\em Differential Inclusions}.
\newblock Springer-Verlag, New York, 1984.

\bibitem{Bin11}
E.~Bini, G.~C. Buttazzo, J.~Eker, S.~Schorr, R.~Guerra, G.~Fohler, K.-E.
  {\AA}rz{\'e}n, R.~Vanessa, and C.~Scordino.
\newblock Resource management on multicore systems: The {ACTORS} approach.
\newblock {\em IEEE Micro}, 31(3):72--81, 2011.

\bibitem{Brecht93}
T.~Brecht.
\newblock On the importance of parallel application placement in {NUMA}
  multiprocessors.
\newblock In {\em Symposium on Experiences with Distributed and Multiprocessor
  Systems (SEDMS IV)}, pages 1--18, San Deigo, CA, July 1993.

\bibitem{Broquedis10}
F.~Broquedis, N.~Furmento, B.~Goglin, P.-A. Wecrenier, and R.~Namyst.
\newblock {ForestGOMP}: An efficient {O}pen{MP} environment for {NUMA}
  architectures.
\newblock {\em Int J Parallel Prog}, 38:418--439, 2010.

\bibitem{chasparis_reinforcement-learning-based_2015}
G.~C. Chasparis.
\newblock Reinforcement-learning-based efficient resource allocation with
  demand-side adjustments.
\newblock In {\em European {C}ontrol {Conference} ({ECC})}, pages 3066--3072,
  Linz, Austria, 2015.

\bibitem{chasparis_stochastic_2019}
G.~C. Chasparis.
\newblock Stochastic stability of perturbed learning automata in
  positive-utility games.
\newblock {\em IEEE Transactions on Automatic Control}, pages 1--1, 2019.

\bibitem{ChasparisAriShamma13_SIAM}
G.~C. Chasparis, A.~Arapostathis, and J.~S. Shamma.
\newblock Aspiration learning in coordination games.
\newblock {\em SIAM J. Control and Optim.}, 51(1), 2013.

\bibitem{chasparis_design_2016}
G.~C. Chasparis, M.~Maggio, E.~Bini, and K.-E. \AA{}rz\'{e}n.
\newblock Design and implementation of distributed resource management for
  time-sensitive applications.
\newblock {\em Automatica}, 64:44--53, February 2016.

\bibitem{chasparis_regression_2016}
G.~C. Chasparis and T.~Natschlaeger.
\newblock Regression {Models} for {Output} {Prediction} of {Thermal} {Dynamics}
  in {Buildings}.
\newblock {\em Journal of Dynamic Systems, Measurement, and Control},
  139(2):021006--021006, November 2016.

\bibitem{epstein_selfish_2008}
L.~Epstein and E.~Kleiman.
\newblock Selfish bin packing.
\newblock In {\em European {Symposium} on {Algorithms}}, pages 368--380.
  Springer, 2008.

\bibitem{Inaltekin05}
H.~Inaltekin and S.~Wicker.
\newblock A one-shot random access game for wireless networks.
\newblock In {\em International Conference on Wireless Networks, Communications
  and Mobile Computing}, 2005.

\bibitem{johansson_randomized_2009}
B.~Johansson, M.~Rabi, and M.~Johansson.
\newblock A randomized incremental subgradient method for distributed
  optimization in networked systems.
\newblock {\em SIAM Journal on Optimization}, 20(3):1157--1170, August 2009.

\bibitem{Khalil92}
H.K. Khalil.
\newblock {\em Nonlinear Systems}.
\newblock Prentice-Hall, 1992.

\bibitem{KushnerYin03}
H.~J. Kushner and G.~G. Yin.
\newblock {\em Stochastic Approximation and Recursive Algorithms and
  Applications}.
\newblock Springer-Verlag New York, Inc., 2nd edition, 2003.

\bibitem{marden_overcoming_2009}
J.~R. Marden and A.~Wierman.
\newblock Overcoming limitations of game-theoretic distributed control.
\newblock In {\em 48th {IEEE} {Conference} on {Decision} and {Control}}, pages
  6466--6471, 2009.

\bibitem{marden_pareto_2014}
J.~R. Marden, H.~P. Young, and L.~Pao.
\newblock Achieving {P}areto optimality through distributed learning.
\newblock {\em SIAM J. Control Optim.}, 52(5):2753--2770, 2014.

\bibitem{Meinhardt99}
H.~Meinhardt.
\newblock Common pool games are convex games.
\newblock {\em J. Public Econ. Theory}, 1(2):247--270, 1999.

\bibitem{nedic_distributed_2009}
A.~Nedic and A.~Ozdaglar.
\newblock Distributed {Subgradient} {Methods} for {Multi}-{Agent}
  {Optimization}.
\newblock {\em IEEE Transactions on Automatic Control}, 54(1):48--61, January
  2009.

\bibitem{Nevelson76}
M.~B. Nevelson and R.~Z. Hasminskii.
\newblock {\em Stochastic Approximation and Recursive Estimation}.
\newblock American Mathematical Society, Providence, RI, 1976.

\bibitem{Reed98}
M.~Reed.
\newblock {\em Fundamental Ideas of Analysis}.
\newblock John Wiley \& Sons, Inc., 1998.

\bibitem{shamma_dynamic_2005}
J.~S. Shamma and G.~Arslan.
\newblock Dynamic fictitious play, dynamic gradient play, and distributed
  convergence to {Nash} equilibria.
\newblock {\em IEEE Transactions on Automatic Control}, 50(3):312--327, March
  2005.

\bibitem{Ste99}
D.~C. Steere, A.~Goel, J.~Gruenberg, D.~McNamee, C.~Pu, and J.~Walpole.
\newblock A feedback-driven proportion allocator for real-rate scheduling.
\newblock In {\em Proc. of the $3^\text{rd}$ Symposium on Operating Systems
  Design and Implementation}, February 1999.

\bibitem{Sub08}
R.~Subrata, A.~Y. Zomaya, and B.~Landfeldt.
\newblock A cooperative game framework for {QoS} guided job allocation schemes
  in grids.
\newblock {\em IEEE Transactions on Computers}, 57(10):1413--1422, October
  2008.

\bibitem{tatarenko_learning_2019}
T.~Tatarenko and M.~Kamgarpour.
\newblock Learning generalized {Nash} equilibria in a class of convex games.
\newblock {\em IEEE Transactions on Automatic Control}, 64(4):1426--1439, April
  2019.

\bibitem{tatarenko_non-convex_2017}
T.~Tatarenko and B.~Touri.
\newblock Non-{Convex} {Distributed} {Optimization}.
\newblock {\em IEEE Transactions on Automatic Control}, 62(8):3744--3757,
  August 2017.

\bibitem{Tembine09}
H.~Tembine, E.~Altman, R.~ElAzouri, and Y.~Hayel.
\newblock Correlated evolutionary stable strategies in random medium access
  control.
\newblock In {\em International Conference on Game Theory for Networks}, pages
  212--221, 2009.

\bibitem{vocking_selfish_2007}
B.~V\"{o}cking.
\newblock Selfish load balancing.
\newblock {\em Algorithmic game theory}, 20:517--542, 2007.

\bibitem{Wei10}
G.~Wei, A.~V. Vasilakos, Y.~Zheng, and N.~Xiong.
\newblock A game-theoretic method of fair resource allocation for cloud
  computing services.
\newblock {\em The Journal of Supercomputing}, 54(2):252--269, November 2010.

\bibitem{yi_initialization-free_2016}
P.~Yi, Y.~Hong, and F.~Liu.
\newblock Initialization-free distributed algorithms for optimal resource
  allocation with feasibility constraints and application to economic dispatch
  of power systems.
\newblock {\em Automatica}, 74:259--269, December 2016.

\bibitem{zhu_approximate_2013}
M.~Zhu and S.~Martinez.
\newblock An {Approximate} {Dual} {Subgradient} {Algorithm} for {Multi}-{Agent}
  {Non}-{Convex} {Optimization}.
\newblock {\em IEEE Transactions on Automatic Control}, 58(6):1534--1539, 2013.

\end{thebibliography}

\end{document}